\documentclass[superscriptaddress,secnumarabic,
amssymb,amsmath,nobibnotes,aps,prd,showkeys,showpacs,nofootinbib]{revtex4-2}%
\usepackage{graphicx}
\usepackage{epsf}
\usepackage{bm}
\usepackage{amsmath}
\usepackage{amsfonts}
\usepackage{amssymb}
\usepackage{epstopdf}
\usepackage{subfigure}
\usepackage{natbib}
\usepackage{color}%
\usepackage{hyperref}

\newtheorem{thm}{Theorem}

\hypersetup{dvips,dvipdfm,linktoc=page,colorlinks=true,linkcolor=blue,citecolor=red,filecolor=magenta,urlcolor=magenta,bookmarks=true}
\providecommand{\U}[1]{\protect\rule{.1in}{.1in}}
\newcommand{\be}{\begin{equation}}
	\newcommand{\ee}{\end{equation}}

\newcommand{\mincir}{\raise
	-3.truept\hbox{\rlap{\hbox{$\sim$}}\raise4.truept\hbox{$<$}\ }}
\newcommand{\magcir}{\raise
	-3.truept\hbox{\rlap{\hbox{$\sim$}}\raise4.truept\hbox{$>$}\ }}

\begin{document}
\title{Interacting Holographic dark energy with matter creation: A dynamical system analysis} 
\author{Goutam Mandal}
\email{gmandal243@gmail.com; 
	rs\_goutamm@nbu.ac.in}
\affiliation{Department of Mathematics, University of North Bengal, Raja Rammohunpur, Darjeeling-734013, West Bengal, India.}
\author{Santosh Biswas}
\email{sant.biswas@gmail.com}
\affiliation{Department of Mathematics, Jadavpur University, Kolkata-700032, West Bengal, India.}
\author{Sujay Kr. Biswas}
\email{sujaymathju@gmail.com; sujay.math@nbu.ac.in}
\affiliation{Department of Mathematics, University of North Bengal, Raja Rammohunpur, Darjeeling-734013, West Bengal, India.}
\begin{abstract}
An interacting Holographic dark energy (HDE) with different infra-red (IR) cutoffs (Hubble horizon and future event horizon) is investigated in the background dynamics of flat Friedmann Lemaitre Robertson Walker (FLRW) universe where gravitational particle creation effects via different form of particle creation rates (1) $\Gamma=3\beta H$ and (2) $\Gamma=3\alpha H_{0}+3\beta H$ are considered. The created particles are considered to be pressureless Dark Matter (DM) which interacts with the HDE through a phenomenological choice of interaction term $Q=3\gamma H \rho_{m}$. We obtain an analytic solution of the cosmological dynamics with Hubble horizon as IR cutoff when the creation rate is taken as $\Gamma=3 \beta H$. On the other hand, employing the Hubble horizon and the future event as IR cutoffs for the model of HDE does not provide the analytic solution when the creation rate is taken as $\Gamma=3\alpha H_{0}+3\beta H$. We then analyze the model separately using the dynamical systems theory. From the analysis, the model (with Hubble horizon as IR cutoff) provides two sets of critical points. One can give a late-time accelerated universe evolving in quintessence, the cosmological constant, or the phantom era. On the other hand, by applying the future event as an IR cutoff, the model provides the complete evolution of the universe. Global dynamics of the model are investigated by defining the appropriate Lyapunov function. The squared sound speed calculated and plotted to find stability of the models.

\end{abstract}
\maketitle
\section{Introduction}
One of the most fascinating issues in modern cosmology is the present acceleration of the expanding universe. This phenomenon is first confirmed by the data collected from Supernovae type I observations \cite{Riess:1998,Perlmutter:1998}. This puzzle cannot be explained theoretically within the General Relativity (GR) by considering that the universe is simply filled with radiation and matter. However, cosmological communities have addressed such issues either by altering the gravity characterised by space-time geometry or by introducing new matter source in the energy-momentum tensor in the Einstein's Field Equations (EFEs). The first alteration is named as `modified gravity theory' based on the modification of space-time geometry of the EFEs. In this context, several modifications have been reported in the literature which includes $f(R)$-gravity \cite{Carroll:2004}, $f(T)$-gravity \cite{Cai:2016}, $f(R,T)$-gravity \cite{Harko:2011},  $f(Q)$-gravity \cite{Lazkoz:2019}, etc. The second approach is the postulating new cosmic fluid with a huge negative pressure, dubbed as Dark Energy (DE) with Equation of State (EoS) parameter satisfying: $\omega_{d}=\frac{p_d}{\rho_{d}}< -\frac{1}{3}$. In this context, the cosmological constant $\Lambda$ can act as the most preferred DE candidate with a constant equation of state parameter (EoS): $\omega_{d}=-1$ for driving the present acceleration. In this context, the cosmological model of $\Lambda$-cold-dark-matter, simply, $\Lambda$CDM, model favored by a huge observational data providing the best fit result upto now. But, this model suffers from two severe theoretical problems such as cosmological constant problem and cosmic coincidence problem \cite{Weinberg:1989,Sahni:2000,Padmanabhan:2003}. To solve these theoretical problems, cosmologists are compelled to look for new models and over the last two decades a numerous DE models having time-varying EoS have been proposed. These are the dynamical DE models with variable equation of state can be capable of providing cosmic speed up. In this regard, the quintessence is the popular and simplest scalar field model which can be able to fulfill the requirement of providing the possible mechanism to address the above problems. There are some other dynamical DE models based on perfect fluid equation of state, have been proposed. Holographic Dark Energy (HDE) model \cite{Li:2004,Wang:2005,Wang:2006,Wang:2017,Horvat:2004,Huang:2004,Pavon:2005,Nojiri:2006,Kim:2006,Setare:2006,Setare:2008,Setare:2009,Ghose:2014,Hsu:2004,Myung:2004} is one of them. The HDE is the theoretical model of DE and is inspired from the famous Holographic Principle (HP) \cite{Hooft:1993,Susskind:1995,Bousso:2002,Fischler:1998,Horava:2000}by applying it to the DE problems. In the HDE model, the DE energy density $\rho_{d}$ only confides on two physical quantities at the boundary of the universe, one is the Plank mass $M_p ^{2}=(8\pi G)^{-1}$, where $G$ is the Newton's gravitational constant and the other is the cosmological length scale $L$ (or sometimes, it is called by Characteristic length). In cosmological context, the $L$ can be particle horizon, event horizon, or the reciprocal of the Hubble parameter $H$.

In the quantum field theory, it is shown (by Cohen et al. \cite{Cohen:1999} ) that the short distance cutoff (i.e., UV cutoff) is related to the long distance cutoff, called Infrared (IR) cutoff due to the limit set by forming a black hole. Then, the quantum zero point energy density, $\rho_{d}$, caused by the short distance cutoff (UV cutoff) will follow the upper limit:
$L^{3} \rho_{d}\leq L M_{p}^{2}$, where $M_p$ is the Plank mass and $L$ is the IR cutoff. That means the total energy of a system of size $L$ cannot exceed the mass of same size ($L$) black hole. Therefore, for a largest $L$, the inequality is saturated and thus, the Holographic energy density is given by:

\begin{equation}\label{HDE}
	\rho_{d}=3b^{2}M_{P}^{2}L^{-2},
\end{equation}
where the $b^{2}$ is a free dimensionless constant and numerical value 3 is used for mathematical convenience. We will set $M_{P}^{2}$ equal to unity, since we shall use the reduced Planck physical units throughout the paper.
The above holographic energy density is comparable to that of the present day dark energy when $L$ is taken as the Hubble scale and equal to the size of the current universe. This concept was discussed in Ref. \cite{Horava:2000}. Note that the cosmological model of holographic dark energy with the Hubble horizon as IR cutoff is not able to describe the present acceleration of the universe \cite{Li:2004,Hsu:2004}. But, the other cutoffs like particle horizon, apparent horizon, future event horizon may describe the present observed acceleration of the universe \cite{Li:2009,Luongo:2017,Setare:2006a,Setare:2007,Setare:2007a,Saridakis:2008,Singh:2019}. Recently, the authors in Ref. \cite{Sadri:2019} showed that an interacting holographic dark energy with Hubble horizon as IR cutoff can give the acceleration within the FRW cosmology. 
Therefore, it is interesting to study the dynamics of an interacting HDE from the Hubble horizon as well as the future event horizon as IR cutoffs. It is worthy to mention that the HDE can be consistent and viable with the current cosmological data as shown by the two recent papers \cite{Tang:2024,Dai:2020}.


Besides these (DE models and modified gravity theories), there is another approach namely, the particle creation mechanism can successfully explain the present acceleration of the universe. Historically, the particle creation mechanism was first introduced by Schrodinger\cite{Schrodinger1939} in a microscopic formalism. Following his idea, creation of quantum particles by gravitational field in curved space-time was discovered in Ref.\cite{Parker Colleraboration}.
Further,  Prigogine et. al. \cite{Prigogine1988,Prigogine1989} studied the macroscopic description of particle creation in the Einstein Field Equations by introducing thermodynamics of open system in cosmology. Subsequently, Calvao et al. \cite{Calvao1992} have extended and generalized the approach in a covariant formulation allowing specific entropy variation as usually expected for non-equilibrium process in fluid. Since then, the cosmological particle creation has been investigated extensively in the literature for describing the cosmic evolution of the universe  \cite{Steigman2009,Lima2010,Lima2014,Fabris2014,Chakraborty2015,Nunes2016} closely mimicking the $\Lambda$CDM model where the created matter mimics as cold dark matter \cite{Prigogine1988,Prigogine1989,Calvao1992,Lima1991,Lima1992,Lima1996,Lima2008,Basilakos2010}. The present acceleration driven by particle creation process is found in Ref. \cite{Nunes2016IJMPD}. In this context, an emergent scenario is reported in Refs. \cite{Chakraborty2014PLBa,Dutta2016} and consequently, a complete cosmic evolution is obtained in Ref. \cite{Chakraborty}.

Thus, cosmological models with the possibilities of particle creation have extensively been studied in various aspects. In one approach, modified gravity theories in the context of adiabatic matter creation process have been studied where the investigators \cite{Harko2014,Harko2015,Harko2021,Pinto2022,Cipriano2024} have considered the non-minimal geometry (curvature, torsion)-matter coupling in an open thermodynamical system and the irreversible energy flow occurred from the gravitational field to created particles. The authors have analysed the cosmological evolution by deriving the particle creation rate, the creation pressure and the entropy production rate directly from the governing equations. 

On the other hand, some other investigators tried to develop the theory of particle production scenario depending upon the choices of particle production rate $\Gamma$ which in general a function of time `t' can only be probed by quantum field theory in curved space-time. However, the quantum field theory can not be able to provide the exact form of this $\Gamma$ and is yet to be developed. Therefore, it is obvious that some phenomenological choices of the form of $\Gamma$ can be made from the cosmological context and the effects of the creation rates are achieved in cosmological evolution. A several choices of $\Gamma$ have been proposed in the literature. 

In particular, the authors in Ref. \cite{Pan2015} have studied $\Gamma\propto H$, $\Gamma\propto \frac{1}{H}$ and have found the matter dominated intermediate phase and transition from deceleration to late time acceleration  phase respectively. Whereas the production rate $\Gamma\propto H^2$ investigated in Ref. \cite{Abramo1996} can provide the early evolution of the universe. On the other hand, investigation of $\Gamma=$ constant in Ref. \cite{Haro} showed the early big-bang singularity to late time de Sitter acceleration. Further, the creation rate is generalized as $\Gamma=\Gamma_{0}+mH+lH^2+\frac{n}{H}$ (see in Ref. \cite{Chakraborty2014PLB})  and is obtained early inflation, late time acceleration and  future deceleration. Furthermore, a more generalization of the creation rate $\Gamma=\Gamma_{0}+\Gamma_{1}H^{-1}+\Gamma_{2}H^{-2}+\sum^{n}_{i=3} \Gamma_{i}H^{-i}$ is found in Ref.\cite{Pan2019}  where a two fluid solutions of particle creation model is performed. They showed in the article that the dynamics of the two fluids is entirely controlled by a single nonlinear differential equation involving the particle creation rate. A singular algebraic solutions of the gravitational field equations are also presented with their stability. Some linear functional forms of creation rate $\Gamma$ are also studied in the cosmological studies. Specifically, $\Gamma=\Gamma_{0}H$ has been studied in Ref. \cite{Biswas2017} in the perspective of dynamical system analysis where the DE is taken as perfect fluid nature and different cosmic phases are found. Further, $\Gamma=3\alpha H_0 +3\beta H$ is considered in Ref. \cite{Singh2020} where the authors have studied the holographic dark energy with the Hubble horizon as well as the event horizon taken as IR cutoffs. Recently, generalised chaplygin gas with baryon has been studied in context of particle creation process with creation rate $\Gamma=3\beta H$ in Ref. \cite{Bhardwaj2024}. Here, a phase transition is shown from deceleration to acceleration phase. It is worthy to note that the authors in Ref. \cite{Nunes2015} studied matter creation model with $\Gamma=3\beta H$ and showed the phantom crossing behavior: $\omega_{eff}<-1$. In a recent paper (see in Ref.\cite{Banerjee2024}) a two-fluid interacting model is studied in context of particle creation mechanism where creation rate is taken as $\Gamma=\Gamma_{0} H_0+\Gamma_{1} H^{-1}$. \\

In this paper we study the cosmological model of HDE when particle creation effect is considered. In the recent past, many authors have investigated particle creation mechanism with interacting two-fluid (DE-DM) model in different aspects. In these studies various form of DE equation of states are assumed. In Ref. \cite{Biswas2017}, interacting DE was investigated in the framework of particle creation scenario where DE was taken as perfect fluid with equation of state $\omega_{d}=\frac{p_d}{\rho_{d}}$ and different interaction terms were taken. Dynamical system analysis was applied in understanding the behaviour of model. In Ref. \cite{Banerjee2024} where an interacting DE-DM cosmological model is investigated in context of particle creation process with creation rate $\Gamma=\Gamma_{0} H_0+\Gamma_{1} H^{-1}$. Dynamical system analysis was performed for a particular form of interaction term. In very recent, an investigation of particle creation effect is also performed through dynamical systems analysis when Umami chaplygin gas with a non-linear equation of state is taken as DE candidate and pressureless dust as dark matter and a complete cosmic evolution with bouncing scenario is obtained (see in Ref.\cite{Mandal:2024}). 
Cosmological models of interacting HDE model with Hubble horizon and the future event horizon as the IR cutoff have been discussed from the dynamical systems perspectives in Refs.\cite{Setare:2009a,Banerjee:2015}. Cosmological emergent scenario is studied in Ref. \cite{Cardenas:2022} where the HDE and cosmological particle creation were taken into account and interaction between HDE and dark matter was not considered.

Holographic dark energy with Hubble horizon as IR cutoff is also investigated in context of particle creation mechanism in Ref. \cite{Singh2020}. The authors in this paper investigated two fluid model without interaction term. They showed the creation rate $\Gamma=3\alpha H_0+3\beta H$ is more effective than that of $\Gamma=0$ or $\Gamma=3\beta H$. The first rate shows the initial matter dominated decelerated phase and transition to accelerated phase. In this paper, we undertake the Hubble horizon and future event horizon as infrared cutoffs to study the interacting model of HDE in context of particle creation with the rates $\Gamma=3\beta H$ and $\Gamma=3\alpha H_0+3\beta H$. We also assume that the dark sectors (HDE and DM) interact through the term $Q=3\gamma H \rho_{m}$ and the possible dynamics is investigated. Main motivation of choosing such complex system is to get an overall idea of the evolution of the universe and behaving as with the observational data. With the first creation rate our model (model 1) of interacting HDE with Hubble horizon as IR cutoff is capable of giving the phase transition for non-zero interaction term ($\gamma\neq 0$). On the other hand, the model is unable to give acceleration for non-interacting case ($\gamma=0$). Next, the second model with creation rate $\Gamma=3\alpha H_0+3\beta H$, we apply dynamical system tools to understand the dynamics of phase space. In the case of Hubble horizon cutoff, the model cannot give the complete cosmic scenario and even matter dominated phase cannot be obtained. However, for the future event horizon, the interacting HDE with the above creation rate provides some insightful results in cosmological context. A complete cosmic scenario is achieved for this case. The universe evolves from early radiation dominated phase to late-time HDE dominated accelerated phase connecting through a matter dominated decelerated intermediate phase. Also, a DM-HDE scaling solution obtained and this can solve the coincidence problem. From the study of global dynamics of the model it is observed that although the scaling solution $P_6$ represents stable locally, but the point $P_2$ is stable globally leading to the late-time HDE dominated accelerated phase is the ultimate phase of the universe.\\

The organization of the paper is as follows:
In the next section \ref{model and autonomous system}, cosmological model of HDE with particle creation is discussed. In section \ref{model 1} we discuss the cosmological dynamics of interacting HDE with particle creation rate $\Gamma=3\beta H$ and an analytic solution is obtained. In section \ref{model 2}, the model of creation rate $\Gamma=3\alpha H_0+3\beta H$ is investigated for interacting HDE by applying Hubble horizon as IR cutoff in subsection \ref{model Hubble horizon} and the future event horizon as IR cutoff is studied in subsection \ref{model event horizon}. The dynamical system analysis is presented to characterise the local stability of the critical points. Adiabatic sound speed is also calculated to derive the model stability for these models of diffrernt IR cutoffs. Global dynamics of the model 2 is presented in subsection \ref{global}. Cosmological implications of the critical points obtained in previous section are discussed in the section \ref{cosmological implications} and overall discussions and concluding remarks are shown in section \ref{conclusions}.



\section{Model with Holographic dark energy and particle creation}\label{model and autonomous system}
Let us consider the universe is homogeneous, isotropic at sufficiently large scale and is characterised by the spatially flat Friedmann-Lemaitre- Robertson-Walker (FLRW) line element:
\begin{equation}\label{FLRW metric}
	ds^2=dt^2 -a^{2}(t) \left(dr^2 +r^2 d\Omega^2\right)
\end{equation}
where $a(t)$ is the scale factor of the universe and  $d\Omega^{2}=d\theta^{2}+sin^{2}\theta d\phi^{2}$ denotes the metric of a 2-dimensional unit sphere.
Within this FLRW cosmology, the matter inside the universe will follow the perfect fluid energy-momentum tensor:
\begin{equation}\label{energy momentum}
	T_{\mu \nu}=(\rho+ P)u_{\mu} u_{\nu}-Pg_{\mu\nu},
\end{equation}
where $u_{\mu}$ is the four-velocity vector of the co-moving fluid satisfying that $u^{\mu}u_{\mu}=1$ and the cosmic substances are represented by their energy density $\rho$ and the thermodynamic pressure $P$. We also consider that the Universe is filled with matter: radiation, dark matter (pressure-less dust) and the dark energy (represented by the HDE). Then the cosmological governing equations is determined by the following Friedmann equation and the acceleration:

\begin{equation}\label{Friedmann}
	3H^{2}=\rho=\sum_{i} \rho_{i}= \rho_{r}+\rho_{m}+\rho_{d},
\end{equation}
\begin{equation}\label{acceleration}
	2\dot{H}+3H^{2}=-P=- \sum_{i} p_{i}= (p_{r}+p_{d})
\end{equation}
 where the subscript $i$ defines the different matters in the universe. Specifically, subscripts $``r"$, $``m"$, and $``d"$ indicate the radiation, dark matter and dark energy respectively. Then the individual components are conserved separately and satisfy their usual continuity equations:
  \begin{equation}\label{radiation}
 	\dot{\rho}_{r}+4H\rho_{r} =0,
 \end{equation}
  
 \begin{equation}\label{DM1}
 	\dot{\rho}_{m}+3H\rho_{m} =0,
 \end{equation}
and
 \begin{equation}\label{DE1}
 	\dot{\rho}_{d}+3H(\rho_{d}+p_{d})=0
 \end{equation} 
  
In the present work irreversible matter creation process is undertaken in an open thermodynamical system where dissipative phenomenon in context of FLRW universe may arise in form of bulk viscous pressure due to non-conservation of particle numbers. In other words, the particles are not covariantly conserved satisfying $N^{\mu}_{;\mu}\neq0$. Here, the universe of a specific volume $V=a^3$ is considered to be containing $N$ number of particles and follow the balance equation:

\begin{equation}\label{balance particle no}
	N^{\mu}_{;\mu}=(nu^{\mu})_{;\mu}= n \Gamma,
\end{equation}
where $n=\frac{N}{V}$ stands for particle number density and $N^{\mu}=n u^{\mu}$ is the particle flow vector and $\dot{n}=n_{\mu} u^{\mu}$. Then, in the background dynamics of FLRW universe, one obtains the following balance equation (from above equation) for particle number density as
\begin{equation}
	\dot{n}+3Hn=n\Gamma,
\end{equation}
Here, $\Gamma$ indicates the matter creation (annihilation) rate which measures the deviation from the standard cosmology. Therefore, the standard cosmology recovers for $\Gamma=0$. However, the effects of non zero rates $\Gamma$ can provide deeper insights in evolution of the cosmological dynamics. 
The above conservation equation for individual dark sector (i.e., dark energy and dark matter) will lead to the following form:
\begin{equation}
	\dot{n}_d+3H n_d=0,
\end{equation}
and 
\begin{equation}
	\dot{n}_m+3H n_m=n_m \Gamma,
\end{equation}
where ($n_d,~ n_m$) are the number density for dark energy and for dark matter respectively. In view of the second law of thermodynamics, the above equation leads to the appearance of negative pressure associated to the creation rate $\Gamma$ and the creation pressure $p_c$ will come into the total energy-momentum tensor with other pressures of matters like dark matter, radiation, dark energy etc. For simplicity, we consider the universe to be open thermodynamical system and the system is isentropic in such a way that the entropy per particle remains conserved, then the creation pressure and the creation rate are related as
\begin{equation}
	p_{c}=-\frac{\Gamma}{3H}(\rho_{m}+p_{m}),
\end{equation}
Here, in cosmological context, one can observe that the cosmic speed up is driven by the above negative pressure. We assume that dark matter particles are created and the produced dark matter is pressure-less dust (i.e., $p_m=0$) and this leads to the following conservation equation for dark matter:
 \begin{equation}\label{DM creation}
	\dot{\rho}_{m}+3H\rho_{m} =\Gamma \rho_{m}.
\end{equation}
Further, a non gravitational interaction is taken into account between the dark sectors, the HDE and the dark matter with $p_m=0$ (i.e., $\omega_{m}=0$). Therefore, they do not conserve separately. By introducing interaction term in (\ref{DM creation}) and (\ref{DE1}), one obtains the continuity equations for the dark matter and the HDE as following: 
\begin{equation}\label{DM interaction}
	\dot{\rho}_{m}+3H\rho_{m}\left( 1-\frac{\Gamma}{3H}\right) =-Q
\end{equation}
and 
\begin{equation}\label{DE interaction}
	\dot{\rho}_{d}+3H(\rho_{d}+p_{d})=Q
\end{equation}
where $Q$ denotes the interaction term and it measures the rate of energy exchange between the dark sectors. In particular, when $Q>0$ the energy exchange occurred from DM to DE, and for $Q<0$, it will be in reverse direction. The total energy density composed by two individual dark sectors (DE and DM) is still conserved:
\begin{equation}\label{total energy}
	\dot{\rho}_{total}+3H(\rho_{total}+p_{total})=0
\end{equation}
where $\rho_{total}=\rho_{d}+\rho_{m}$ and $p_{total}=p_d +p_c$.
However, from many observations, it is speculated that the dark sectors (DE and DM) are the dominant source of the present universe and an interaction between them can play an important role in solving cosmological issues. An appropriate  interacting model is capable of providing a possible mechanism to alleviate the coincidence problem by allowing the energy exchange between the dark sectors. Thus, an interacting model which provides the scaling attractor having similar order of energy densities $$\frac{\Omega_{d}}{\Omega_{m}}\approx \mathcal{O} (1)$$ and with $q<0$ or $\omega_{eff}<-\frac{1}{3}$, can fit the observation and can alleviate the coincidence problem by choices of the model parameters without fine tuning the initial conditions. In this context, one may follow the Ref. \cite{Boehmer 2008} where the authors have studied the interacting dark energy and have presented the issue clearly in perspective of dynamical system analysis. Interested readers may also follow some other references (see for instance \cite{Biswas:2015,Odintsov2018a,Odintsov2018b,Mishra:2019a,Mishra:2019b,Aljaf2020,Biswas:2020}) for studying dark energy models with the help of dynamical system analysis. For more details, we may cite the review works in Ref. \cite{Bahamonde2018}. It is worthy to mention that some recent works on cosmological dynamical systems could be considered for the reference list, for example in Ref. \cite{Oikonomou:2024} it was shown for the first time that coupled DM scalar systems may lead to Gravitational Waves (GW)  production, and also phase space analysis of invariant spaces was executed in Ref. \cite{Chatzarakis:2020}.

It is interesting to note that interacting DE model can also provide the possible solution of recent $H_0$ tension. In this context, the reader can follow the reference \cite{Pan:2023} and the references therein. 
However, there is no guiding principle to choice the interaction term. One may choose it phenomenologically and for mathematical simplicity. The authors in Ref. \cite{Zimdahl:2005} showed a unified picture of different interacting HDE models at the perturbative level. In the present work, we consider the interaction term $Q\propto H \rho_{m}$ from phenomenological point of view and for the mathematical simplicity. The explicit form of interaction is the following \cite{Zimdahl:2001,Zimdahl:2003,Wang:2005}:
\begin{equation}\label{interaction}
	Q=3\gamma H \rho_{m}
\end{equation}
Where $\gamma$ stands for dimensionless coupling constant which measures the strength of interaction between the dark sectors and the numerical value 3 is used for mathematical convenience. For $\gamma>0$ the energy flow occurs from DM to DE and the flow will be in reverse direction for $\gamma<0$. The non-interacting case is for $\gamma=0$. Several forms of particle creation rates ($\Gamma$) are considered in studying Cosmological dynamics with particle creation mechanism. We will consider a simple form of creation rate $\Gamma$ for investigating the dynamics of interacting HDE with pressure-less dust. In the next section, we shall study the interacting HDE with different Infrared cutoff in context of particle creation with different creation rates.

%



\section{Model 1: Interacting HDE with creation rate $\Gamma=3\beta H$}\label{model 1}
In this section, we consider the HDE and employ the Hubble horizon as an IR cutoff for which the cosmological length scale $L$ is taken as $L=H^{-1}$, where $H$ denotes the Hubble parameter. Considering the above expression of $L$ and by using the Planck units, the Eqn. (\ref{HDE}) gives the explicit form of HDE energy density as

\begin{equation}\label{Hubble horizon IR}
	\rho_{d}=3b^{2}H^{2}.
\end{equation}
Then, the equation of state parameter for the HDE reads as
\begin{equation}\label{Hubble p}
	p_{d}=\omega_{d}\rho_{d}.
\end{equation}
In this study, we also adopt the particle creation rate: 
\begin{equation}\label{creation rate 1}
	\Gamma=3 \beta H
\end{equation}
where $\beta$ is a dimensionless constant. Using the interaction term ( $Q=3 \gamma H \rho_{m}$ ) in Eqn (\ref{interaction}) and the creation rate $\Gamma=3 \beta H$ in Eqn (\ref{creation rate 1}), the conservation equation in  Eqn (\ref{DM1}) will give the energy density for DM:
\begin{equation}\label{energy_DM}
	\rho_{m}=\rho_{m0} a^{-3(1+\gamma-\beta)}
\end{equation}
and the energy conservation equation for radiation (\ref{radiation}) will provide the energy density for radiation as 
\begin{equation}\label{energy_radiation}
	\rho_{r}=\rho_{r0} a^{-4}
\end{equation}
 where $\rho_{m0}$ and $\rho_{r0}$ are current energy densities for DM and radiation respectively.\\
\begin{figure}
	\centering
	\subfigure[]{%
		\includegraphics[width=8.5cm,height=8.5cm]{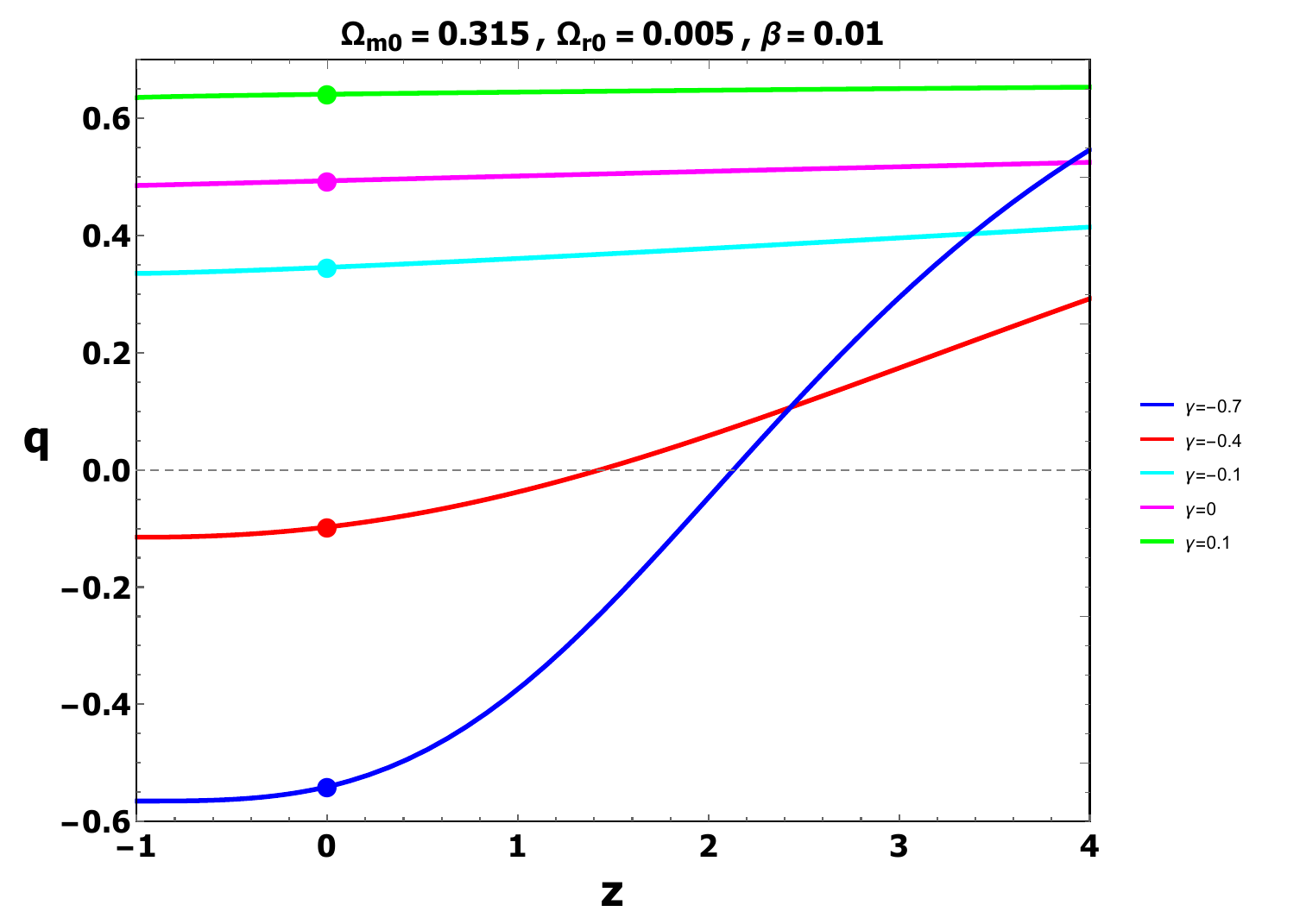}\label{Hubble_q}}
	\qquad
	\subfigure[]{%
		\includegraphics[width=8.5cm,height=8.5cm]{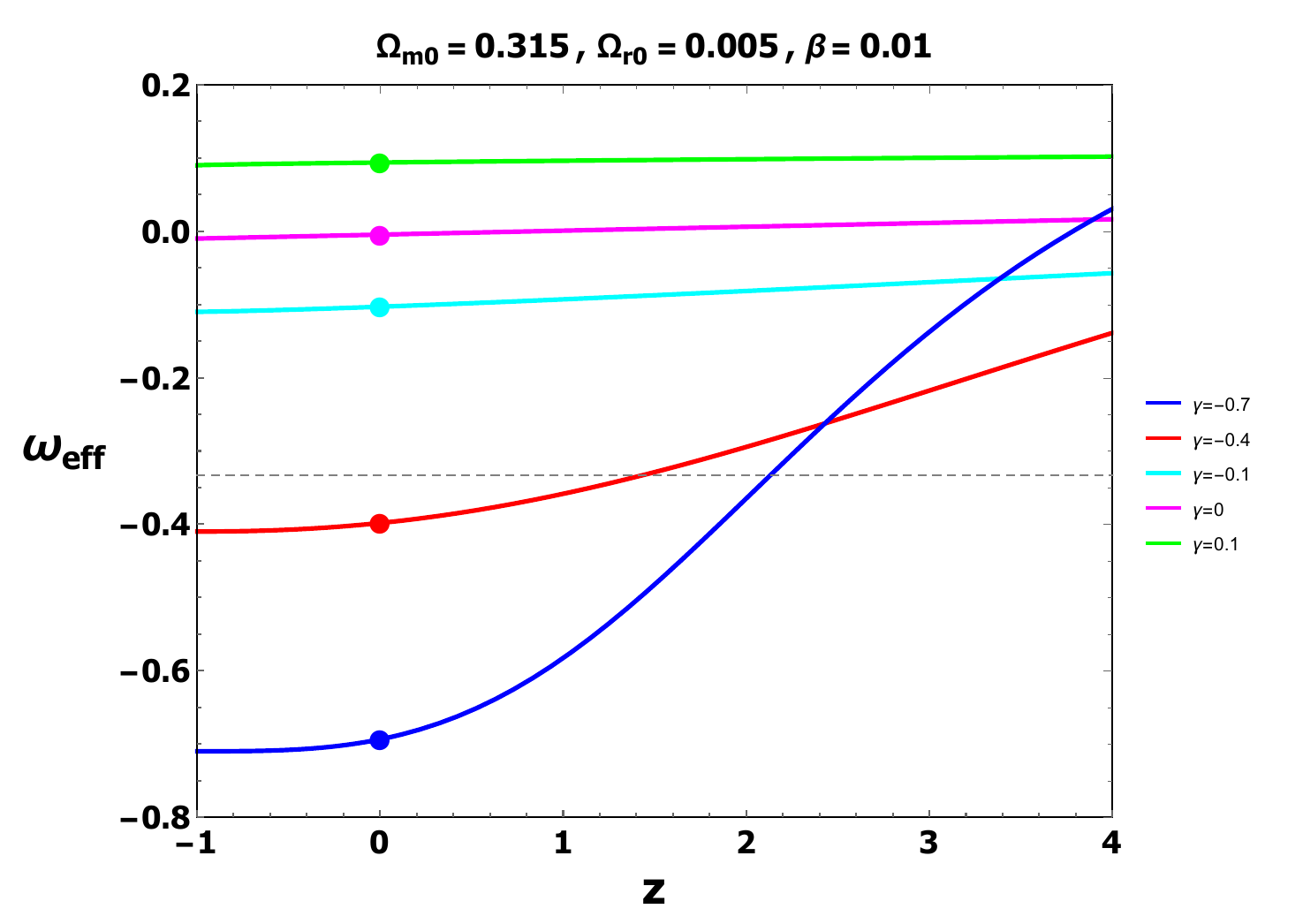}\label{Hubble_omega_eff}}
	\qquad
	\subfigure[]{%
		\includegraphics[width=8.5cm,height=8.5cm]{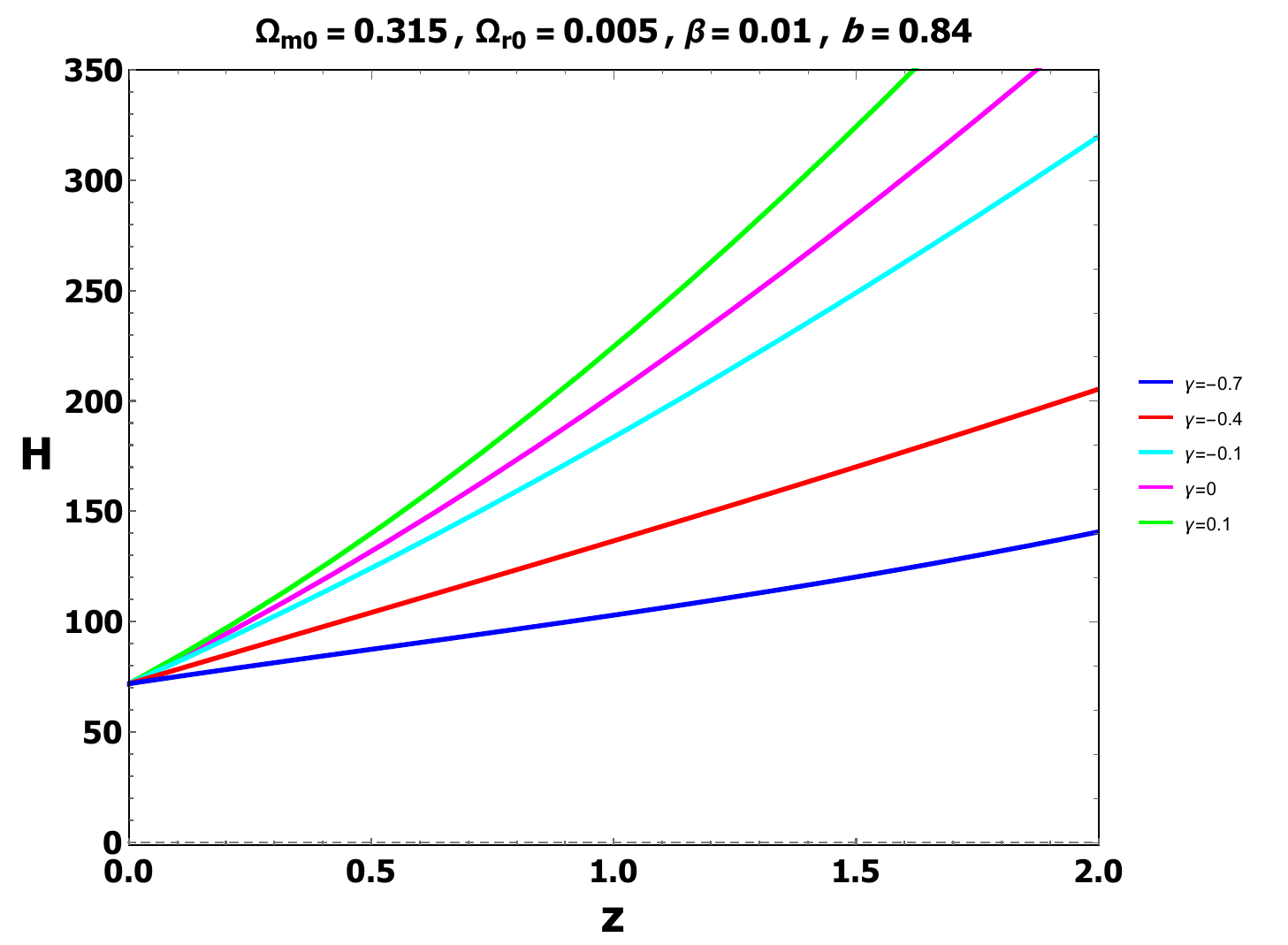}\label{Hubble_H}}
	\caption{ Figure shows the behaviour of deceleration parameter ($q$), effective equation of state parameter ($\omega_{eff}$) and Hubble parameter ($H$) against the redshift $z$ in \ref{Hubble_q}, \ref{Hubble_omega_eff} and \ref{Hubble_H} respectively for different values of coupling parameter $\gamma$. In sub-figure \ref{Hubble_H} the present value of the Hubble parameter is $H_0=69 ~~kms^{-1} Mpc^{-1}$ \cite{Pawde}.}
	\label{fig 1}
\end{figure} 

Now, using expressions of energy densities of DM, radiation and HDE in Eqns (\ref{energy_DM}),  (\ref{energy_radiation}) and (\ref{Hubble horizon IR}), the Friedmann equation (\ref{Friedmann}) provides the expression of Hubble parameter in terms of redshift ($z$) as:
 \begin{equation}
 	E(z)=\frac{H(z)}{H_0}=\left[\frac{1}{1-b^2}\left\lbrace \Omega_{m0}(1+z)^{3(1+\gamma-\beta)} +\Omega_{r0}(1+z)^4  \right\rbrace  \right]^\frac{1}{2}.
 \end{equation}
By using the above equation one can get easily (after some calculations) the deceleration parameter in terms of redshift in the following form
\begin{equation}
	q=-1+(1+z)\frac{E_{z}(z)}{E(z)}=\frac{\Omega_{m0} (1-3 \beta +3 \gamma ) (z+1)^{3 \gamma }+2 \Omega_{r0} (1+z)^{1+3 \beta }}{2 \left\lbrace \Omega_{m0} (1+z)^{3 \gamma }+\Omega_{r0} (1+z)^{1+3 \beta}\right\rbrace }
\end{equation}
and the effective equation of state parameter in terms of redshift
\begin{equation}
	\omega_{eff}=\frac{3 \Omega_{m0} (\gamma -\beta ) (1+z)^{3 \gamma }+\Omega_{r0} (1+z)^{1+3 \beta }}{3 \Omega_{m0} (1+z)^{3 \gamma }+3 \Omega_{r0} (1+z)^{1+3 \beta}}
\end{equation}
where $\Omega_{m0}$, $\Omega_{r0}$ are current fractional density parameters for DM and radiation respectively.
In fig. (\ref{fig 1}), we show the evolution of the deceleration parameter ($q$), effective equation of state parameter $\omega_{eff}$ and the Hubble parameter $H$ with respect to redshift $z$ for different values of interaction coupling parameter $\gamma$. The sub-figure \ref{Hubble_q}  and sub-fig \ref{Hubble_omega_eff} show that for non-interacting case ($\gamma=0$), there is no phase transition from deceleration to acceleration as it is shown in Ref. \cite{Singh2020}, while non-zero  interaction ($\gamma\neq 0$) is allowed the model will be able to give the phase transition. 



\section{Model 2: Interacting HDE with creation rate $\Gamma=3\alpha H_{0}+3\beta H$}\label{model 2}
In this section, we shall analyse model of HDE interacting with pressureless dust through an interaction term referred in Eqn (\ref{interaction}) in context of gravitational particle production. The creation rate is assumed as linear proportional to the current value of the Hubble parameter ($H_{0}$) and the Hubble function($H$) 
\begin{equation}\label{creation rate 2}
	\Gamma=3\alpha H_{0}+3\beta H
\end{equation} 
where the parameter $\alpha$, $\beta$ lie in the interval $[0,1]$. We adopt Hubble horizon and future event horizon as IR cutoffs for the HDE to study the model in two different subsections.

\subsection{HDE with Hubble horizon as IR cutoff}\label{model Hubble horizon}
First we adopt Hubble horizon as IR cutoff which yields the holographic energy density as of the form in Eqn (\ref{Hubble horizon IR}). Field equations with the interaction in Eqn (\ref{interaction}) and the creation rate in Eqn (\ref{creation rate 2}) are complicated in form and the exact analytical solution of the model cannot be extracted. So, dynamical analysis of this model is executed to get an overall idea of evolution of the universe.
\subsubsection{ Dynamical analysis: Phase space variables, autonomous system and critical points}\label{sec-Autonomous}

Here, we shall apply dynamical system tools and techniques to the model because the evolution  equations are non-linear and complicated in form. We constitute  autonomous system of Ordinary Differential Equations (ODEs) by proper choices of variables. We consider below the dimensionless dynamical variables in terms of cosmological variables as
\begin{equation}\label{variables}
	x=\frac{\rho_{m}}{3H^{2}},~~u=\frac{p_{d}}{3H^{2}},~~\mbox{and}~~z=\frac{H_{0}}{H+H_{0}}.
\end{equation} 
which are normalized over Hubble scale. 
Note that the fractional contribution of the radiation energy density $\Omega_{r}$, the dark matter energy density $\Omega_{m}$ and the dark energy density $\Omega_{d}$ will not be able to form a compact phase space, we therefore, introduce a new dimensionless variable 
$z=\frac{H_{0}}{H+H_{0}}$ as a dynamical variable satisfying $0\leq z \leq 1$. This type of variable is recently used in Ref. \cite{Halder:2024} to study a compact phase space of interacting DE model. Note that  $H=\infty \Longrightarrow z=0$ and $H=0$ corresponds to $z=1$ which may be a singularity in cosmological evolution as well as dynamical evolution of this model. Then the evolution equations of variables are in the form of autonomous system of ordinary differential equations:
\begin{eqnarray}\label{autonomous}
	\begin{split}
		\frac{dx}{dN}& =x\left\lbrace -3\gamma-b^{2}+\left(1+3\beta+\frac{3\alpha z}{1-z} \right)(1-x)+3u \right\rbrace ,& \\
		\frac{du}{dN}& =-\frac{b^{2}}{3(1-b^{2})}\left[x\left\lbrace -3\gamma-b^{2}+\left(1+3\beta+\frac{3\alpha z}{1-z} \right)(1-x)+3u \right\rbrace\left(1+3\beta-\frac{3\gamma}{b^{2}}+\frac{3\alpha z}{1-z} \right) -\frac{3\alpha x}{(1-z)^{2}} \right]  ,& \\
		\frac{dz}{dN}& = \frac{z(1-z)}{2}\left\lbrace 4-b^{2}+3u-x\left( 1+3\beta+\frac{3\alpha z}{1-z} \right) \right\rbrace .
		&
	\end{split}
\end{eqnarray}
Here, $N = \ln a$ is taken as independent variable, called the e-folding number. The system is singular at $z=1$ i.e., at $H=0$. To study of this system we would like to avoid such singularity. We do not undertake any critical point at $z=1$. We are interested only in the late time dynamics of the model. To do so, first we express all the physical parameters to be expressed in terms of dynamical variables $x$, $u$ and $z$ (the co-ordinates of critical points):\\
The fractional contribution of density parameter for the HDE is:
\begin{equation}\label{density parameter DE}
	\Omega_{d}=b^{2}.
\end{equation}
The density parameter for dark matter can take the form
\begin{equation}\label{density parameter DM}
	\Omega_{m}=x,
\end{equation}
The density parameter for radiation can take the value
\begin{equation}\label{density parameter radiation}
	\Omega_{r}=1-x-b^{2}.
\end{equation}
The equation of state parameter for HDE
reads as:
\begin{equation}\label{EOS DE}
	\omega_{d}=\frac{p_{d}}{\rho_{d}}=\frac{1}{3}\left\lbrace 1-b^{2}+3u-x\left( 1+3\beta-\frac{3\gamma}{b^{2}}+\frac{3\alpha z}{1-z} \right) \right\rbrace.
\end{equation}
The deceleration parameter can be expressed in the form
\begin{equation}\label{deceleration}
	q=1+\frac{1}{2} \left\lbrace b^2 \left(3 u-b^2\right)+3 \gamma  x-x\left(1+b^2\right) \left(1+3 \beta +\frac{3 \alpha  z}{1-z}\right)\right\rbrace 
\end{equation}
and the effective equation of state parameter will be of the form
\begin{equation}\label{EOS effective}
\omega_{eff}=\frac{1}{3} \left\lbrace1+ b^2 \left(3 u-b^2\right)+3 \gamma  x-x\left(1+b^2\right) \left(1+3 \beta +\frac{3 \alpha  z}{1-z}\right)\right\rbrace. 
\end{equation}
Moreover, we have the evolution equation of the Hubble function as
\begin{equation}\label{evolution Hubble}
	-\frac{2\dot{H}}{H^{2}}=4-b^{2}+3u-x\left( 1+3\beta+\frac{3\alpha z}{1-z} \right) 
\end{equation}
Now, condition for the acceleration is $q<0$ or $\omega_{eff}<-\frac{1}{3}$. 

Phase space boundary is obtained as
\begin{equation}
\Psi_{\mbox{s1}}=\left\{   0\leq x\leq 1,~~u\in R,~~0\leq z \leq 1.  \right \}
\end{equation}

In this section, We determine the critical points of the above autonomous system and then we perturb the equations up to first order about the critical points, in order to determine their stability. we present the existence of the critical points and the corresponding physical parameters, eigenvalues in detail in tabular form.\\
The critical points for this system (\ref{autonomous}) are the following:
{\bf 
	\begin{itemize}
		\item  I. Set of Critical Points : $C_{1}=(0,u_{c},0)$
		\item  II. Set of Critical Points : $ C_{2}=(0,\frac{1}{3} \left(b^2-4\right),z_{c})$
	\end{itemize}
	
}
The existence of the sets of critical points and their cosmological parameters are displayed in the table (\ref{physical_parameters}). Also, the eigenvalues corresponding to the set of critical points are presented in the table (\ref{eigenvalues})

	\begin{table}[tbp] \centering
		\caption{The existence of the sets of critical points and the corresponding physical parameters for the system (\ref{autonomous}) with interaction $Q=3\gamma H \rho_{m}$ and creation rate $\Gamma=3\alpha H_{0}+3\beta H$. }%
		\begin{tabular}
			[c]{cccccccccc}\hline\hline
			\textbf{Critical Points}&$\mathbf{\Omega_{r}}$&$\mathbf{\Omega_{m}}$& $\mathbf{\Omega_{d}}$ & $\mathbf{\omega_{d}}$ &
			$\mathbf{\omega_{eff}}$ &  $q$ & Existence &
			\\\hline
			$C_{1}$ & $1-b^{2}$  & $0$ & $b^{2}$ &
			$\frac{1}{3} \left(1-b^2+3 u_{c}\right)$ & $\frac{1}{3} \left\lbrace 1+b^2 \left(3 u_{c}-b^2\right)\right\rbrace $ & $1+\frac{1}{2} b^2 \left(3 u_{c}-b^2\right)$  & $\forall \alpha, \beta, \gamma~\mbox{and}~0\leq b^{2}\leq 1$ \\ \\
			$C_{2}  $ & $1-b^{2}$ & $0$ & $b^{2}$ &
			$-1$ & $\frac{1}{3} \left(1-4 b^2\right)$ & $1-2 b^2$  & $\forall \alpha,\beta, \gamma~\mbox{and}~0\leq b^{2}\leq 1$ 
				\\\hline\hline
		\end{tabular}
		\label{physical_parameters} \\
		
\end{table}%
%


\begin{table}[h!] \centering
	\caption{The eigenvalues corresponding to the sets of critical points of the system (\ref{autonomous}) are presented}%
	\begin{tabular}
		[c]{cccccccc}\hline\hline
		\textbf{Critical Point}&$\mathbf{\lambda_1}$& $\mathbf{\lambda_2}$ & $\mathbf{\lambda_3}$ & 
		\\\hline
		$C_1  $ & $0$ & $\frac{1}{2} \left(4-b^2+3 u_{c}\right)$ & $1-b^2+3 \beta -3 \gamma +3 u_{c}$
		\\ \\
		$C_2  $ & $0$ & $0$ & $\frac{3 \left\lbrace \beta +z_{c} (\alpha -\beta +\gamma +1)-\gamma -1\right\rbrace }{1-z_{c}}$ &
	   \\
		\\\hline\hline
	\end{tabular}
	\label{eigenvalues} \\
	
\end{table}%
%
\begin{figure}
	\centering
	\subfigure[]{%
		\includegraphics[width=7.2cm,height=7.2cm]{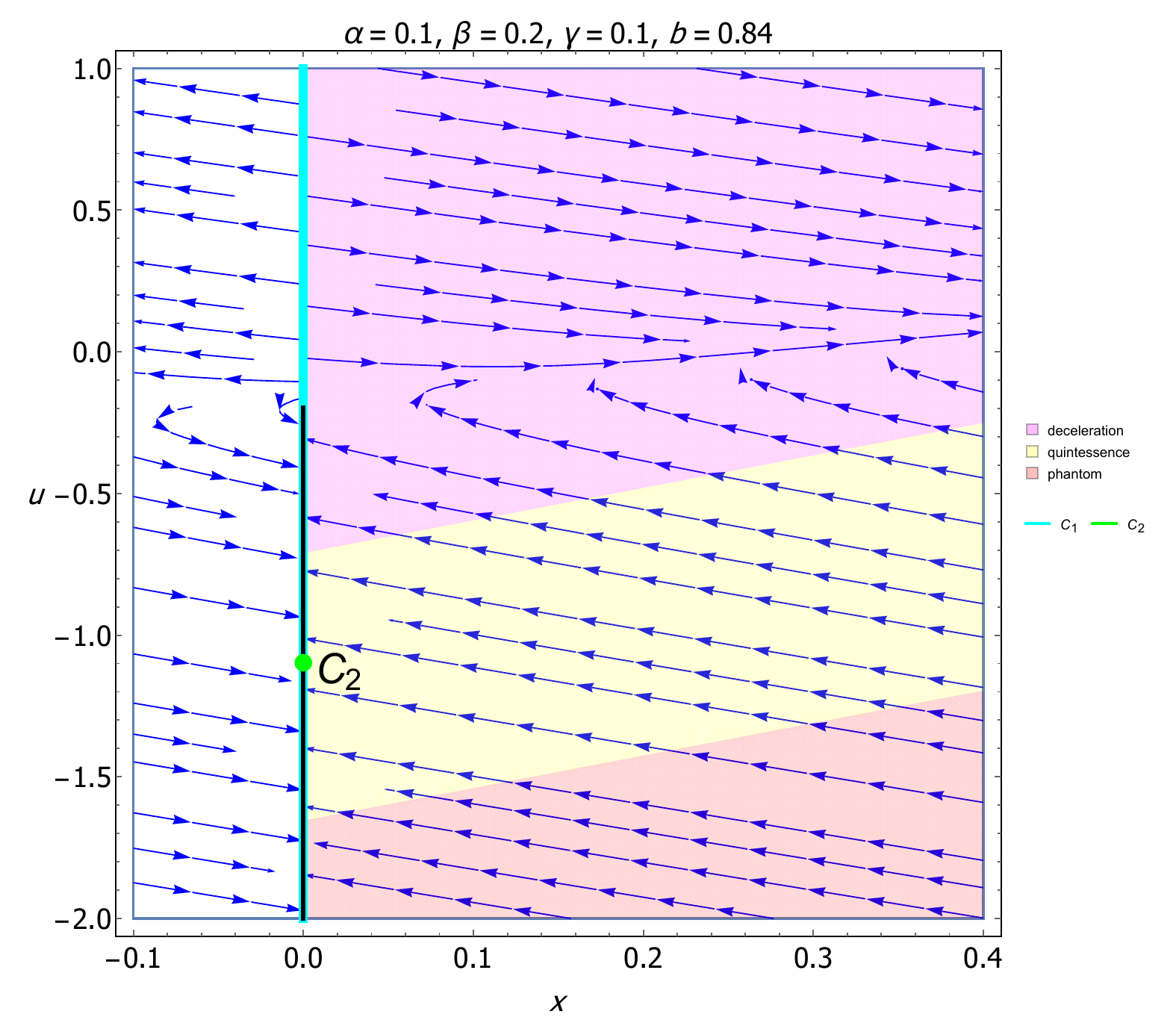}\label{Hubble_xu}}
	\qquad
	\subfigure[]{%
		\includegraphics[width=7.2cm,height=7.2cm]{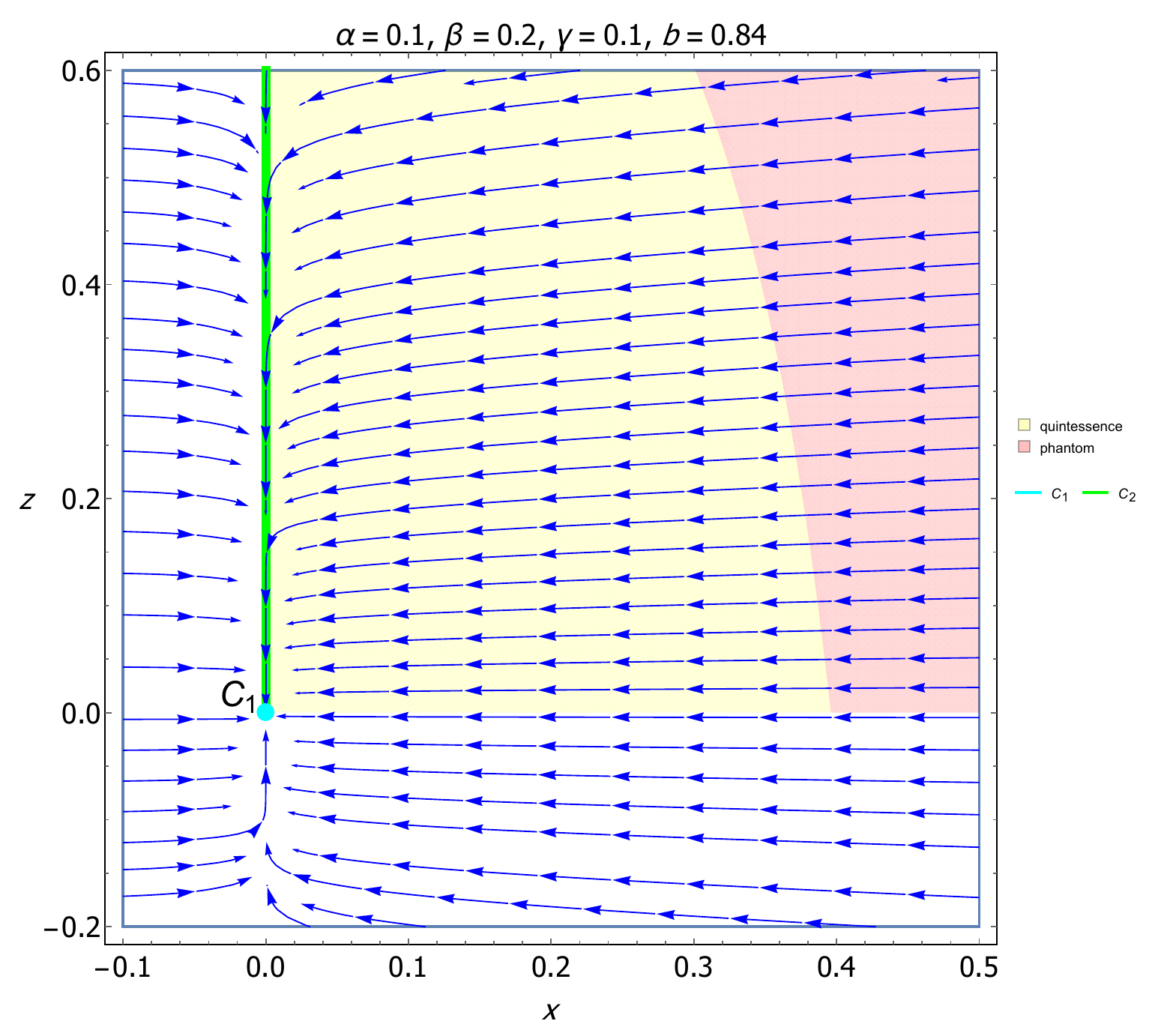}\label{Hubble_xz}}
	\qquad
	\subfigure[]{%
		\includegraphics[width=7.2cm,height=7.2cm]{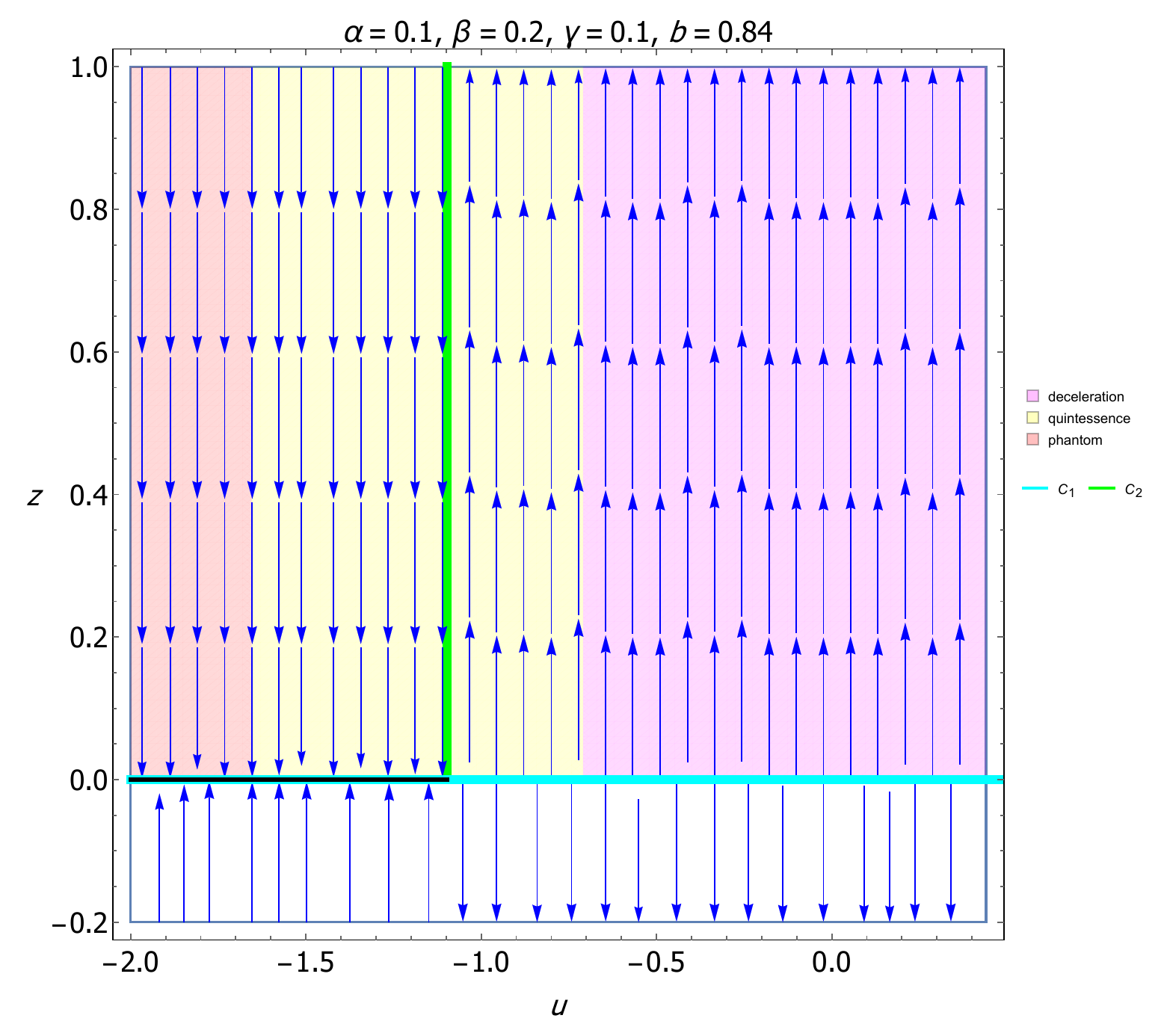}\label{Hubble_uz}}
	\caption{ The phase space projections for the autonomous system (\ref{autonomous}) are plotted on different planes for parameter values $\alpha=0.1,~~\beta=0.2,~~\gamma=0.1,~~b=0.84$. Sub-figures \ref{Hubble_xu} and \ref{Hubble_uz} show that the point $C_1$ is DE dominated accelerated solution for $u_c<-0.7096$ and point $C_2$ is DE dominated accelerated saddle-like solution. Here, the yellow-shaded region represents the quintessence region (i.e. $-1<\omega_{eff} < -\frac{1}{3}$), the red-shaded region represents the phantom region (i.e. $\omega_{eff} < -1$) and the magenta-shaded region corresponds to the decelerated region (i.e. $\omega_{eff} \geq -\frac{1}{3}$). The cyan color line represents the set of critical points $C_1$ and the green color line represents the set $C_2$. The black lines for $u_c<-0.1981$ in \ref{Hubble_xu} and for $u_c<-1.0981$ in \ref{Hubble_uz} correspond to the stable points on the set $C_1$. }
	\label{Point C1 C2}
\end{figure} 

 \subsubsection{Local stability of critical points: Phase space analysis }\label{sec-phase space analysis}
 In this section, we shall investigate the local stability of the present model.
 
 \begin{itemize}
 	\item The set of critical points $C_1$ exists for all free parameters ($\alpha$, $ \beta$, $\gamma$, $b$) satisfying $0\leq b^{2}\leq 1$ and $u_c \in R$ in the phase space $x-u-z$. The fractional density parameters for dark energy, dark matter and radiation $(\Omega_{d},~\Omega_{m},~\Omega_{r})=(b^2,~0,~1-b^2)$ indicate that the set of points $C_1$ corresponds to solution with combination of both DE and radiation. Dark matter is absent here. 
 	Points on the set will become completely DE dominated when $b\approx 1$ ($\Omega_{r}\approx0,~\Omega_{m}=0,~\Omega_{d}\approx1$ see table \ref{physical_parameters}). On the other hand, the set will become in complete radiation dominated for $b\approx 0$ ($\Omega_{r}\approx1,~\Omega_{m}=0,~\Omega_{d}\approx0$). The HDE corresponding to the set has the perfect fluid equation of state $\omega_{d}=\frac{1}{3} \left(1-b^2+3 u_{c}\right)$. Therefore, the HDE can have different natures in its evolution. The behaviour of quintessence like fluid is observed for $b^2 -4<3u_c <b^2 -2$. On the other hand, it behaves as cosmological constant like fluid for $3u_c =b^2 -4$ and as phantom like fluid is achieved by the HDE for $3u_c <b^2 -4$. The accelerated expansion is viewed by the set of points for $q\equiv 1+\frac{1}{2} b^2 \left(3 u_{c}-b^2\right)<0$, otherwise, it is decelerated. The condition for acceleration is: ($q<0$, $\omega_{eff}<-\frac{1}{3}$) :\
 	$u_{c}<-\frac{1}{3}~~\mbox{and}~~ \left(-1\leq b<-\sqrt{\frac{1}{2} \sqrt{9 u_{c}^2+8}+\frac{3 u_{c}}{2}}~~\mbox{or}~~ \sqrt{\frac{1}{2} \sqrt{9 u_{c}^2+8}+\frac{3 u_{c}}{2}}<b\leq 1\right)$.\\
 The stability of the set $C_1$ can be found by evaluating the eigenvalues of linearised Jacobi matrix. It has two dimensional non-empty stable sub-manifold corresponding to the negative eigendirections i.e.,  for $\lambda_{2}<0$ and $\lambda_{3}<0$.
 		
However, the set $C_{1}$ is a non-isolated set of critical points and it has exactly one vanishing eigenvalue. Therefore, it is a normally hyperbolic set, and the stability of this type of set can be found by observing the signature of the remaining non-vanishing eigenvalues. In the case of the set $C_{1}$, the conditions for the stability are the following:
 	
 			$b^2+3 \gamma >1+3 \beta +3 u_{c}~~  \mbox{and}~~\\
 			(i)~~\left\lbrace -\frac{4}{3}\leq u_{c}<-1~~\mbox{and}~~ \left(-1\leq b<-\sqrt{3 u_{c}+4},~~\mbox{or}~~\sqrt{3 u_{c}+4}<b\leq 1\right)\right\rbrace ,~~\mbox{or}~~ (ii)~~-1\leq b\leq 1~~\mbox{and}~~ u_{c}<-\frac{4}{3} $ \\
On the other hand, the set $C_{1}$ will be unstable (source) solution when the following conditions are satisfied:\\
 		$b^2+3 \gamma <1+3 \beta +3 u_{c}~~\mbox{and}\\
 		(i)~~( u_{c}\leq -1~~\mbox{and}~~ b+\sqrt{3 u_{c}+4}>0~~\mbox{and}~~ \sqrt{3 u_{c}+4}>b),~~\mbox{or}~~
 		(ii)~~ u_{c}>-1~~\mbox{and}~~ -1\leq b\leq 1$\\
Finally, the set $C_1$ will have the saddle like nature for the following conditions:\\
 		 $b^2+3 \gamma >1+3 \beta +3 u_{c}~~\mbox{and}\\
 		 (i)~~(u_{c}\leq -1~~\mbox{and}~~ b+\sqrt{3 u_{c}+4}>0~~\mbox{and}~~ \sqrt{3 u_{c}+4}>b),~~\mbox{or}~~
 		 (ii)~~u_{c}>-1~~\mbox{and}~~ -1\leq b\leq 1$\\
From the above analysis, one can classify that the set of points $C_1$ can evolve in different eras at late-times. The late time accelerated era of the universe is achieved by the set when it evolves in quintessence era (i.e. $\lambda_2<0, \lambda_3<0~\mbox{and}~ -1<\omega_{eff}<-\frac{1}{3}$) for  ~
 		  $b^2+3 \gamma >1+3 \beta +3 u_{c}~~\mbox{and}\\
 		  (i)~ \left\lbrace -\frac{7}{6}<u_{c}<-1~~\mbox{and}~~ \left(\sqrt{3 u_{c}+4}<b<\sqrt{\frac{1}{2} \sqrt{9 u_{c}^2+16}+\frac{3 u_{c}}{2}}~~\mbox{or}~~ -\frac{\sqrt{\sqrt{9 u_{c}^2+16}+3 u_{c}}}{\sqrt{2}}<b<-\sqrt{3 u_{c}+4}\right)\right\rbrace ,~~\mbox{or}~~\\
 		  (ii)~ \left\lbrace u_{c}\leq -\frac{7}{6}~~\mbox{and}~~ \left(\sqrt{\frac{1}{2} \sqrt{9 u_{c}^2+8}+\frac{3 u_{c}}{2}}<b<\sqrt{\frac{1}{2} \sqrt{9 u_{c}^2+16}+\frac{3 u_{c}}{2}}~~\mbox{or}~~ -\frac{\sqrt{\sqrt{9 u_{c}^2+16}+3 u_{c}}}{\sqrt{2}}<b<-\frac{\sqrt{\sqrt{9 u_{c}^2+8}+3 u_{c}}}{\sqrt{2}}\right)\right\rbrace $
On the other hand, the set represent the late time accelerated evolution of the universe attracted in cosmological constant era (i.e. $\lambda_2<0, \lambda_3<0~\mbox{and}~ \omega_{eff}=-1$) for ~
 		 $u_{c}<-1~~\mbox{and}\\
 		 (i)~ \left\lbrace b=-\sqrt{\frac{1}{2} \sqrt{9 u_{c}^2+16}+\frac{3 u_{c}}{2}}~~\mbox{and}~~ \gamma >\frac{1}{3} \left(1-b^2+3 \beta +3 u_{c}\right)\right\rbrace ,~~\mbox{or}~~\\
 		 (ii)~ \left\lbrace b=\sqrt{\frac{1}{2} \sqrt{9 u_{c}^2+16}+\frac{3 u_{c}}{2}}~~\mbox{and}~~ \gamma >\frac{1}{3} \left(1-b^2+3 \beta +3 u_{c}\right)\right\rbrace $.\\
Also, the set $C_1$ will represent the late time accelerated evolution of the universe evolving in phantom regime (i.e. $\lambda_2<0, \lambda_3<0~\mbox{and}~ \omega_{eff}<-1$) for ~
 		  $u_{c}<-1~~\mbox{and}\\
 		  (i)~\left\lbrace -1\leq b<-\sqrt{\frac{1}{2} \sqrt{9 u_{c}^2+16}+\frac{3 u_{c}}{2}}~~\mbox{and}~~ \gamma >\frac{1}{3} \left(1-b^2+3 \beta +3 u_{c}\right)\right\rbrace ,~~\mbox{or}~~\\
 		  (ii) ~\left\lbrace \sqrt{\frac{1}{2} \sqrt{9 u_{c}^2+16}+\frac{3 u_{c}}{2}}<b\leq 1~~\mbox{and}~~ \gamma >\frac{1}{3} \left(1-b^2+3 \beta +3 u_{c}\right)\right\rbrace $.\\
 		  
Thus, one may conclude that the set of critical points $C_1$ is normally hyperbolic set and it provides some interesting solutions in late-time cosmology. It shows present acceleration of the universe evolving in quintessence era, cosmological constant era or in phantom regime. 		  
It is also to be noted that the set can exhibit the decelerated unstable source (i.e. $\lambda_2>0, \lambda_3>0~\mbox{and}~ \omega_{eff}>-\frac{1}{3}$) for the following conditions:\\ ~
 		   $b^2+3 \gamma <3 \beta +3 u_{c}+1~~\mbox{and}\\
 		   (i)~ \left(u_{c}\leq -\frac{7}{6}~~\mbox{and}~~ b+\sqrt{3 u_{c}+4}>0~~\mbox{and}~~ \sqrt{3 u_{c}+4}>b\right),~~\mbox{or}~~\\
 		   (ii)~ \left(-\frac{7}{6}<u_{c}\leq -\frac{1}{3}~~\mbox{and}~~ -\frac{\sqrt{\sqrt{9 u_{c}^2+8}+3 u_{c}}}{\sqrt{2}}<b<\sqrt{\frac{1}{2} \sqrt{9 u_{c}^2+8}+\frac{3 u_{c}}{2}}\right),~~\mbox{or}~~\\
 		   (iii)~ \left(u_{c}>-\frac{1}{3}~~\mbox{and}~~ -1\leq b\leq 1\right)$.
\begin{figure}
	\centering
	\subfigure[]{%
		\includegraphics[width=5.4cm,height=5.4cm]{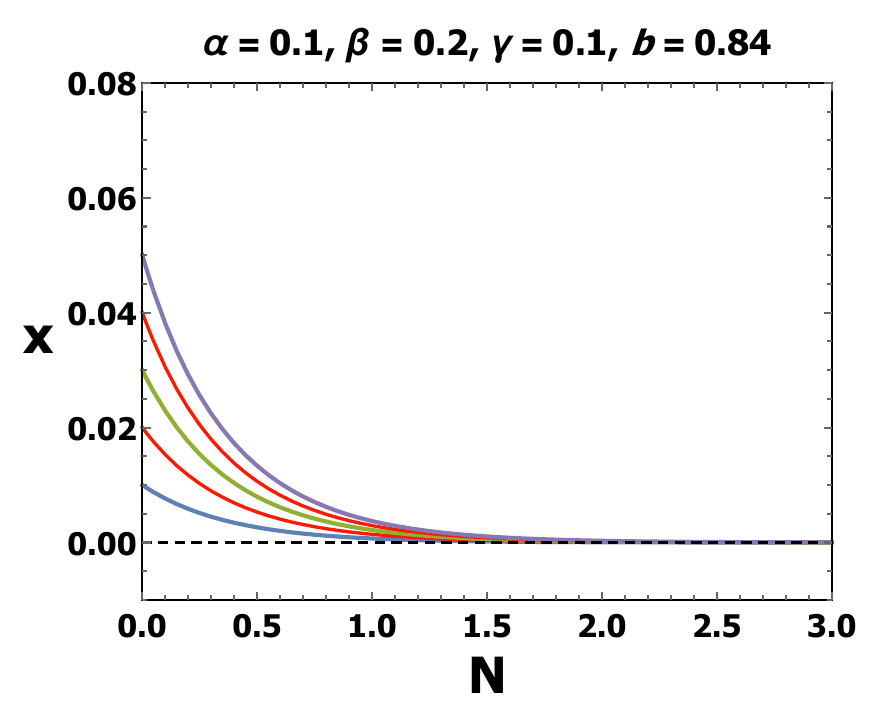}\label{Hubble_Nx}}
	\qquad
	\subfigure[]{%
		\includegraphics[width=5.4cm,height=5.4cm]{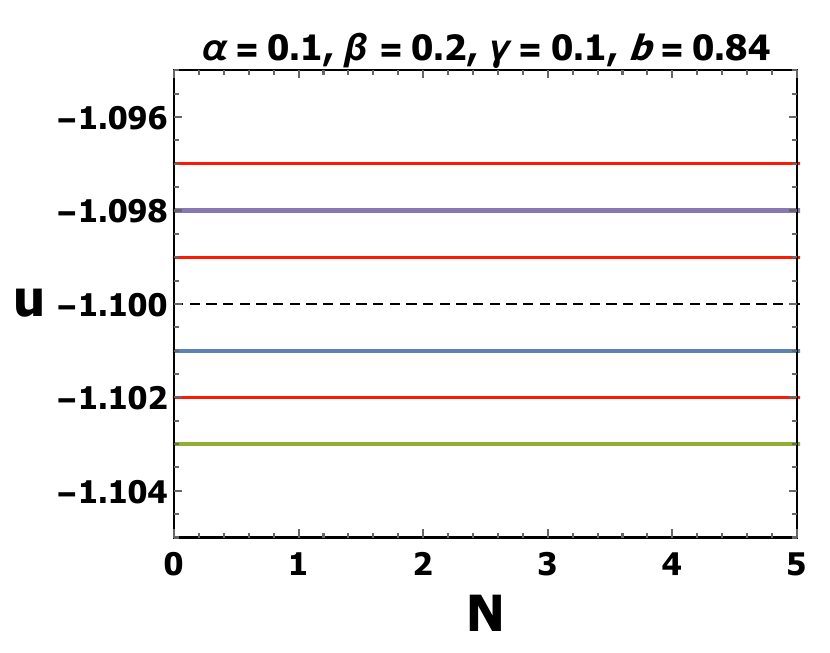}\label{Hubble_Nu}}
	\qquad
	\subfigure[]{%
		\includegraphics[width=5.4cm,height=5.4cm]{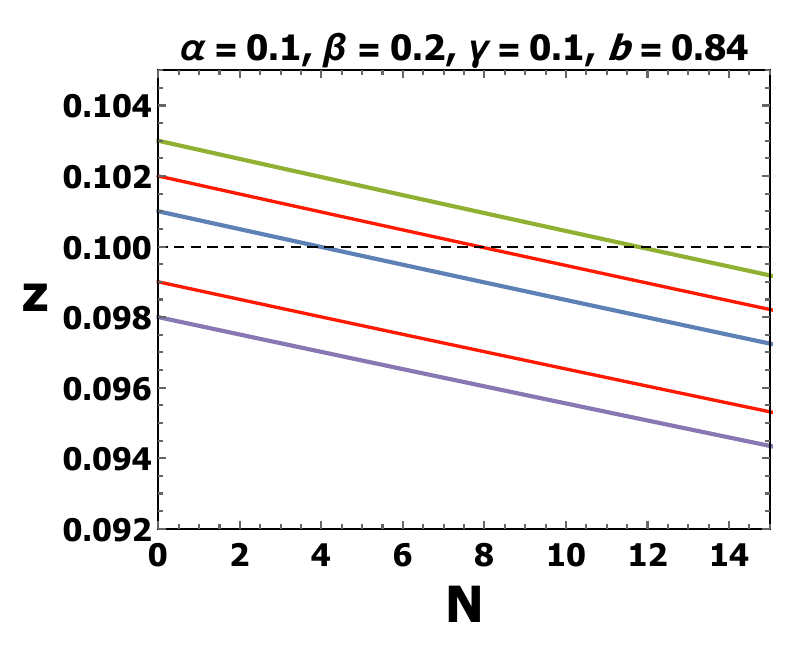}\label{Hubble_Nz}}
	\caption{ The figure shows the time evolution of trajectories of $x(N)$, $u(N)$ and $z(N)$ for the critical point $C_{2}$ for parameter values $\alpha=0.1$, $\beta=0.2$, $\gamma=0.1$ and $b=0.84$. Sub- fig. \ref{Hubble_Nx} exhibits that the perturbations come back but sub-figures \ref{Hubble_Nu}, \ref{Hubble_Nz} show that perturbations do not come which leads to unstable nature of the set of points $C_2$. }
	\label{C2 unstable}
\end{figure}
\begin{figure}
	\centering
	\includegraphics[width=11.6cm,height=9.3cm]{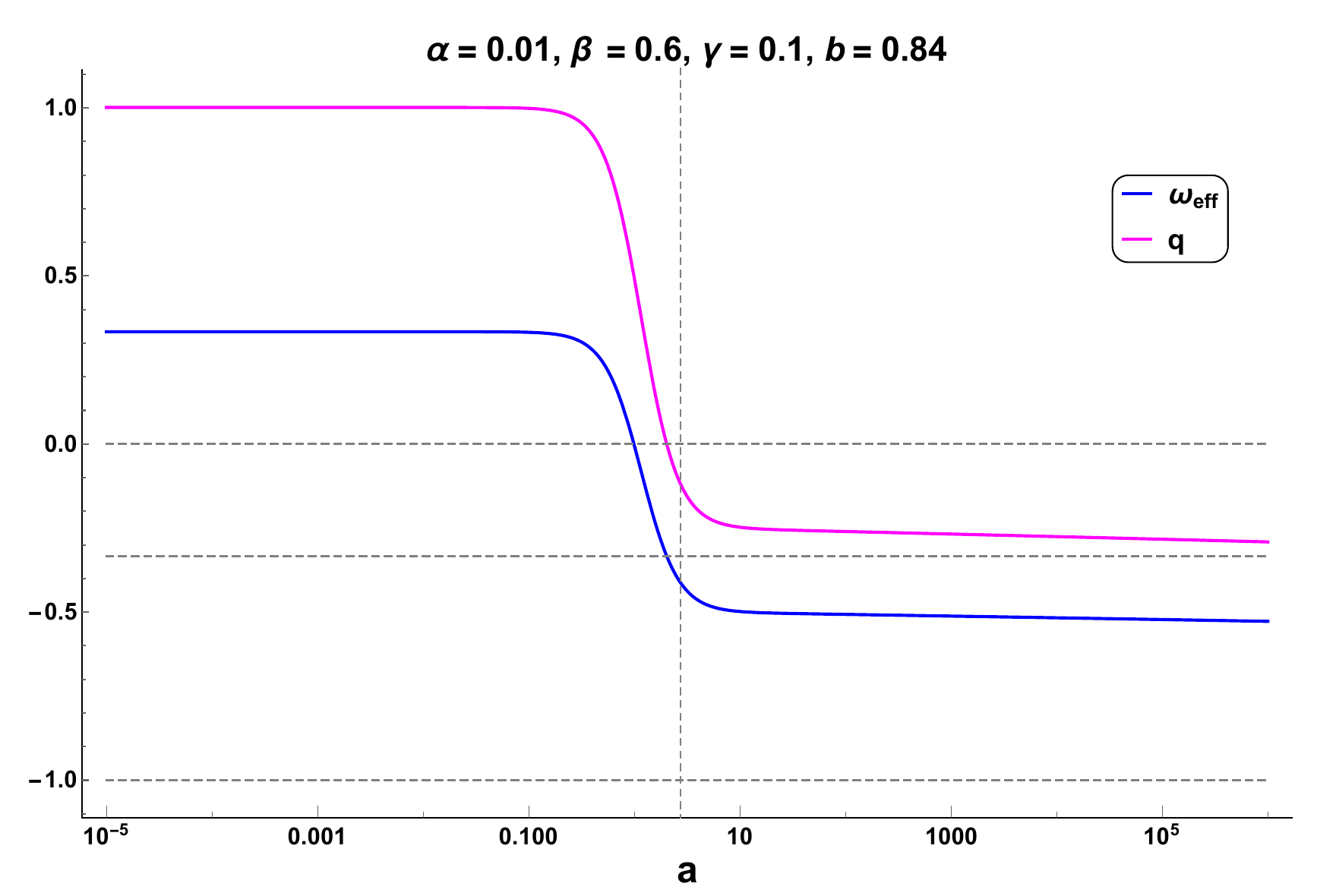}
	
	\caption{ The figure shows the evolution of cosmological parameters: effective equation of state ($\omega_{eff}$) and the deceleration parameter ($q$) for the free parameters values: $(\alpha,~\beta,~\gamma,~ b)=(0.01,~0.6,~0.1,~0.84)$. There exists phase transition from deceleration to the acceleration era of the universe.
	}
	\label{Hubble_evolution}
\end{figure} 

 \item The set of critical points $C_2$ exists for all free parameters $\alpha$, $ \beta$, $\gamma$ and $0\leq b^{2}\leq 1$ and $0\leq z_c\leq 1 $ in the phase space $x-u-z$. The set of points $C_2$ is the solution with combination of HDE and radiation only and DM is absent ($\Omega_{r}=1-b^2$, ${\Omega_{m}}=0$ and ${\Omega_{d}}=b^2$). Points on the set will become completely DE dominated for $b=1$ ($\Omega_{r}=0,~\Omega_{m}=0,~\Omega_{d}=1$ see table \ref{physical_parameters}). HDE for this set will always behave as a cosmological constant like fluid since $\omega_{d}=-1$. Acceleration will exists for $q<0\Longrightarrow$ $b^2>\frac{1}{2}$.\\
 The set of critical points $C_{2}$ is non hyperbolic in nature (since it has two vanishing eigenvalues) and it has one dimensional stable sub-manifold in the negative eigendirection corresponding to $\lambda_{3}<0$ and it has one dimensional unstable sub-manifold for $\lambda_{3}>0$. However, the stability of the set is found numerically. The analysis shows that the set is unstable in nature.\\
 However, the set $C_{2}$ evolves in quintessence era (i.e. $-1<\omega_{eff}<-\frac{1}{3}$) for:\\
 		   $\left( -1<b<-\frac{1}{\sqrt{2}}~~\mbox{or}~~ \frac{1}{\sqrt{2}}<b<1\right) $\\
and it evolves in cosmological constant (i.e. $\omega_{eff}=-1$) era for:\\
 		   $\left( b=-1~~\mbox{or}~~ b=1\right) $\\
For both the scenarios, the accelerating nature is achieved for the set. However, from numerical analysis (see in fig. \ref{C2 unstable}, the plot shows the set is unstable), it is observed that the evolution cannot be a late-time scenario.		   
 		   
It is also noted that the set cannot cross the Phantom barrier (i.e. $\omega_{eff}<-1$).
 		  
\end{itemize}

Note: The dynamical system (\ref{autonomous}) has only one stable critical point $C_1$ and remaining one critical point $C_2$ is unstable in nature. Then the  stability of unique  critical point $C_1$ of dynamical system (\ref{autonomous}) implies its global stability. As its domain of attraction is the entire space in $R^3$.\\

\subsubsection{Classical stability of the model}

The stability of the model is characterised by the square of the sound speed ($C_{s}^{2}$). This has a crucial role in assessing the model whether it is stable or unstable. The model is assumed to be stable for square of sound speed is greater than zero, i.e., $C_{s}^{2}>0$ and when it is less than zero, then it is deemed to be unstable i.e., for $C_{s}^{2}<0$. The expression for the sound speed can be calculated by
\begin{equation}\label{sound speed formula}
	C_{s}^{2}=\frac{\delta p(t)}{\delta \rho(t)}~.
\end{equation}
By using the Eqns. (\ref{DE1}), (\ref{Hubble horizon IR}), (\ref{variables}), and (\ref{evolution Hubble}) in Eqn. (\ref{Hubble p}), we express explicit form of the sound speed in (\ref{sound speed formula}) in terms of the dynamical variables $x(N),~u(N),~z(N)$ which is given in the following for the model of Hubble horizon is taken as IR cutoff,
 	\begin{align*}
 	C_{s}^{2}=\frac{1}{3} \Bigg[\frac{(z-1) \left\lbrace x \left(1+3 \beta-\frac{3 \gamma }{b^2} -\frac{3 \alpha  z}{z-1}\right) \left(3u-b^2-3 \gamma +(1-x) \left(1+3 \beta -\frac{3 \alpha  z}{z-1}\right)\right)-\frac{3 \alpha  x}{(z-1)^2}\right\rbrace }{\left(b^2-1\right) \left\lbrace b^2 (z-1)+x (-3 \alpha  z+3 \beta  (z-1)+z-1)-(3 u+4) (z-1)\right\rbrace }+x \left(\frac{3 \gamma }{b^2}-3 \beta +\frac{3 \alpha  z}{z-1}-1\right) \\
 	-b^2+3 u+1 \Bigg] 
 \end{align*}	   
 	  \begin{figure}
 	  	\centering
 	  	\includegraphics[width=16cm,height=9cm]{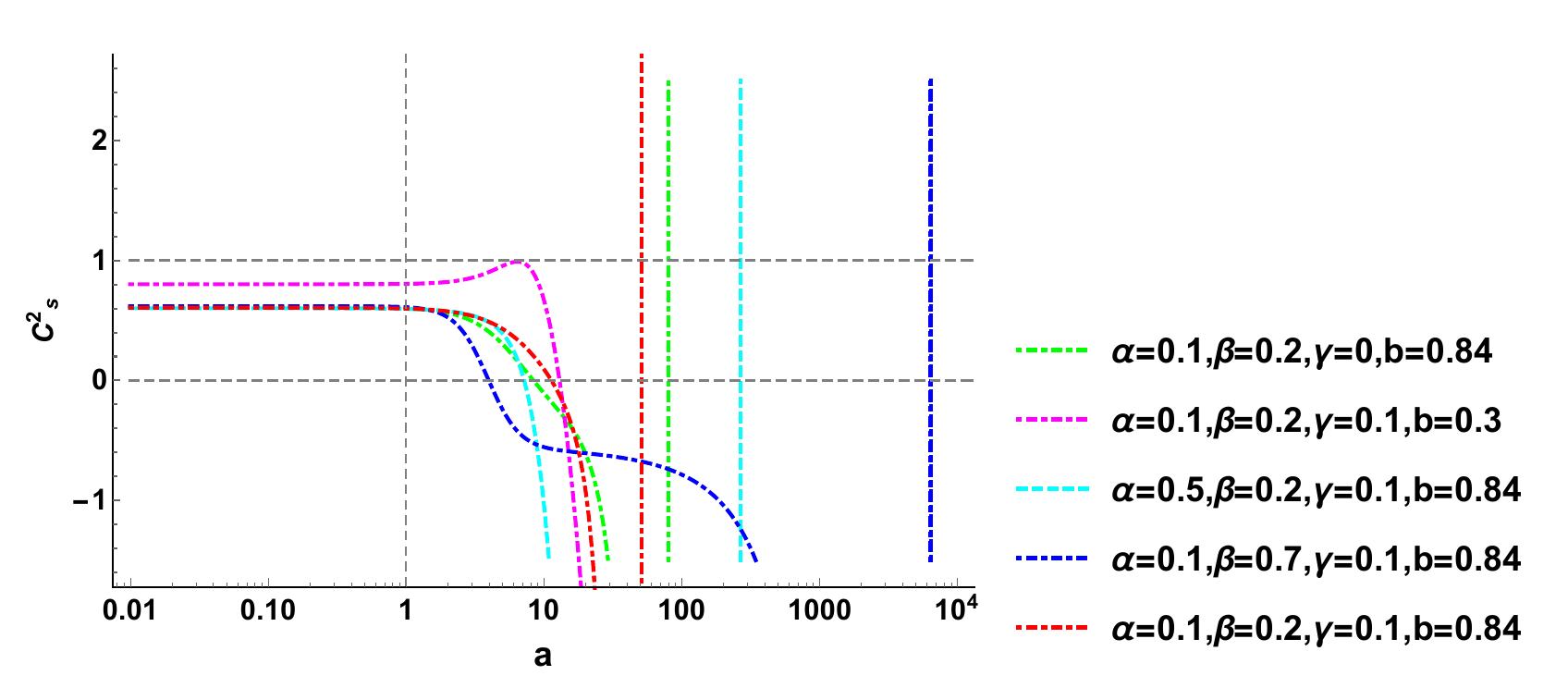}
 	  	
 	  	\caption{The figure shows sound speed curves of HDE with Hubble horizon as IR cutoff for different choices of model parameters.
 	  	}
 	  	\label{sound_evolution}
 	  \end{figure} 
The fig. (\ref{sound_evolution}) is displayed the evolution of the square of speed of sound ($C_{s}^{2}$) for different model parameters. The figure exhibits that the model is stable during its early epoch and will become unstable (later) in future epoch. So, it can be concluded that the model is unstable since the curves of $C_{s}^{2}$ is not always positive ($C_{s}^{2}>0$) during the whole evolutionary epoch.

\subsection{HDE with future event horizon as IR cutoff}\label{model event horizon} 	   
 The future event horizon is defined as,
 \begin{equation}
 	R_{h}=a \int_{t}^{\infty}\frac{dt}{a}=a \int_{a}^{\infty}\frac{da}{a^2H}~.
 \end{equation}
 Differentiating (\ref{HDE}) and using (\ref{DE1}) we get,
 \begin{equation}\label{event energy conservation}
 	3H\rho_{d}(1+\omega_{d})-Q=2\dot{R_{h}}R_{h}^{-1}\rho_{d},
 \end{equation}
 where, $\dot{R_{h}}=HR-1$.
 Taking $R_h=\frac{b}{H\sqrt{\Omega_d}}$,  (\ref{event energy conservation}) we have,
 \begin{equation}\label{Event omega}
 	\omega_{d}=-\frac{1}{9}\left(\frac{6\sqrt{\Omega_d}}{b}-\frac{Q}{\Omega_d H^3}+3 \right). 
 \end{equation}

 
 \subsubsection{Dynamical analysis: Phase space variables, autonomous system and critical points}\label{sec-Autonomous1}
 
 As the evolution  equations are non-linear and complicated in form, exact analytical solutions cannot be obtained. Therefore, we shall apply dynamical system tools and techniques to extract the information of evolution from the model because this analysis allows us to bypass the non-linearities and complications of cosmological evolution equations. For that one has to convert the evolution equations into an autonomous system of Ordinary Differential equations (ODEs) by proper transformation of variables. We consider below the dynamical variables in terms of cosmological variables as follows :
 \begin{equation}\label{variables1}
 	x=\frac{\rho_{m}}{3H^{2}},~~y=\frac{\rho_{d}}{3H^{2}},~~\mbox{and}~~z=\frac{H_{0}}{H+H_{0}}.
 \end{equation} 
 which are normalized over Hubble scale. 
 Now, choosing the particle production rate $\Gamma$ as a linear combination of the current value of the Hubble parameter ($H_{0}$) and the Hubble function($H$) as in Eqn \ref{creation rate 2} and the interaction $Q=3\gamma H\rho_{m}$ in Eqn \ref{interaction}, where the coupling parameter $\gamma$ is a dimensionless constant, the autonomous system of ordinary differential equations are obtained as:
 \begin{eqnarray}\label{autonomous1}
 	\begin{split}
 		\frac{dx}{dN}& =x\left\lbrace \left(1+3\beta-3\gamma+\frac{3\alpha z}{1-z} \right)(1-x)-2y\left(1+\frac{\sqrt{y}}{b} \right)  \right\rbrace ,& \\
 		\frac{dy}{dN}& =y\left\lbrace2(1-y)\left(1+\frac{\sqrt{y}}{b} \right)  -x\left(1+3\beta-3\gamma+\frac{3\alpha z}{1-z} \right) \right\rbrace  ,& \\
 		\frac{dz}{dN}& = \frac{z(1-z)}{2}\left\lbrace 4-2y\left(1+\frac{\sqrt{y}}{b} \right)  -x\left(1+3\beta-3\gamma+\frac{3\alpha z}{1-z} \right) \right\rbrace  .
 		&
 	\end{split}
 \end{eqnarray}
 Here, the independent variable is chosen as the lapse time $N = \ln a$, which is called the e-folding number.
 Now, physical parameters can be expressed in terms of dynamical variables $x$, $y$ and $z$ (the co-ordinates of critical points) as follows:
 the density parameter for dark energy i.e, for Holographic dark energy(HDE) takes the value:
 \begin{equation}\label{density parameter DE1}
 	\Omega_{d}=y,
 \end{equation}
 The density parameter for dark matter can take the form
 \begin{equation}\label{density parameter DM1}
 	\Omega_{m}=x,
 \end{equation}
 The density parameter for radiation can take the value
 \begin{equation}\label{density parameter radiation1}
 	\Omega_{r}=1-x-y,
 \end{equation}
 The equation of state parameter for HDE
 reads as:
 \begin{equation}\label{EOS DE1}
 	\omega_{d}=\frac{p_{d}}{\rho_{d}}=-\frac{1}{3}\left(1+\frac{2\sqrt{y}}{b}-\frac{3\gamma x}{y} \right) 
 \end{equation}
 [which is a non-constant parameter leading to the condition
 to be a dark energy $\omega_{d}<0.$]
 The deceleration parameter can be expressed in the form
 \begin{equation}\label{deceleration}
 	q=1-y\left(1+\frac{\sqrt{y}}{b} \right)-\frac{x}{2}\left(1+3\beta-3\gamma+\frac{3\alpha z}{1-z} \right)  
 \end{equation}
 and the effective equation of state parameter will be of the form
 \begin{equation}\label{EOS effective}
 	\omega_{eff}=\frac{1}{3} \left\lbrace 1-2y\left(1+\frac{\sqrt{y}}{b} \right)-x\left(1+3\beta-3\gamma+\frac{3\alpha z}{1-z} \right) \right\rbrace .
 \end{equation}
 Moreover, we have the evolution equation of the Hubble function as
 \begin{equation}\label{evolution Event}
 	-\frac{2\dot{H}}{H^{2}}=4-2y\left(1+\frac{\sqrt{y}}{b} \right)-x\left(1+3\beta-3\gamma+\frac{3\alpha z}{1-z} \right)
 \end{equation}
 Now, for acceleration either $q<0$ or $\omega_{eff}<-\frac{1}{3}$. For the energy condition, the
 phase space becomes bounded in physical region of dynamical variables as
 \begin{equation}
 \Psi_{\mbox{s2}} =  \left\{  0\leq x\leq 1,~~0\leq y \leq 1,~~0\leq z \leq 1.   \right\}	
 \end{equation}
 
 In this section, We determine the critical points of the above autonomous system and then we perturb the equations up to first order about the critical points, in order to determine their stability. we present the critical points and the corresponding physical parameters, eigenvalues in detail in tabular form.\\
 The critical points for this system (\ref{autonomous1}) are the following:
 {\bf 
 	\begin{itemize}
 		\item  I. Critical Point : $P_{1}=(0,0,0)$
 		\item  II. Critical Point : $ P_{2}=(0,1,0)$
 		\item  III. Critical Point : $ P_{3}=(0,b^2,0)$
 		\item  IV. Critical Point : $ P_{4}=(1,0,0)$
 		\item  V. Critical Point : $ P_{5}=(1,0,\frac{1-\beta+\gamma }{1+\alpha -\beta +\gamma })$
 		\item  II. Critical Point : $ P_{6}=(\frac{1}{4} \left(4-b^2 (1-3 \beta +3 \gamma )^2\right),\frac{1}{4} b^2 (1-3 \beta +3 \gamma )^2,0)$
 		\item  II. Critical Point : $ P_{7}=(1-b^2,b^2,\frac{1-\beta+\gamma }{1+\alpha -\beta +\gamma })$
 	\end{itemize}
 	
 }
 The critical points and their cosmological parameters of the system (\ref{autonomous1}) are displayed in the table (\ref{physical_parameters1}). Also, the eigenvalues corresponding to the critical points of the system (\ref{autonomous1}) are presented in the table (\ref{eigenvalues1})

 
 	\begin{table}[tbp] \centering
 		\caption{The critical points and the corresponding physical parameters for the system (\ref{autonomous1}) with interaction $Q=3\gamma H \rho_{m}$ and creation rate $\Gamma=3\alpha H_{0}+3\beta H$, where $\Delta=1-3 \beta +3 \gamma$ }%
 		\begin{tabular}
 			[c]{cccccccccc}\hline\hline
 			\textbf{Critical Points}&$\mathbf{\Omega_{r}}$&$\mathbf{\Omega_{m}}$& $\mathbf{\Omega_{d}}$ & $\mathbf{\omega_{d}}$ &
 			$\mathbf{\omega_{eff}}$ &  $q$ & 
 			\\\hline
 			$P_{1}$ & $1$  & $0$ & $0$ &
 			$\frac{0}{0}$ & $\frac{1}{3} $ & $1$  \\ \\
 			$P_{2}  $ & $0$ & $0$ & $1$ & $-\frac{b+2}{3 b}$ & $-\frac{b+2}{3 b}$ & $-\frac{1}{b}$ \\ \\
 			$P_{3}  $ & $1-b^2$ & $0$ & $b^2$ & $-1$ & $\frac{1}{3} \left(1-4 b^2\right)$ & $1-2 b^2$\\ \\
 			$P_{4}  $ & $0$ & $1$ & $0$ & $\frac{1}{0}$ & $\gamma -\beta$ & $1-\frac{3}{2} \left(\beta -\gamma +\frac{1}{3}\right)$ \\ \\
 			$P_{5}  $ & $0$ & $1$ & $0$ & $\frac{1}{0}$& $-1$ & $-1$  \\ \\
 			$P_{6}  $ & $0$ & $\frac{1}{4} \left(4-b^2 \Delta^2\right)$ & $\frac{1}{4} b^2 \Delta^2$ & $\frac{\gamma  \left(4-b^2 \Delta^2\right)}{b^2 \Delta^2}+\beta -\gamma -\frac{2}{3}$ & $\frac{1}{6} \left(6\gamma-b^2 \Delta^3-6 \beta \right)$ & $\frac{1}{4} \Delta \left(2-b^2 \Delta^2\right)$ \\ \\
 			$P_{7}  $ & $0$ & $1-b^2$ & $b^2$ & $\left(\frac{1}{b^2}-1\right) \gamma -1$ & $-1$ & $-1$ 
 			\\\hline\hline
 		\end{tabular}
 		\label{physical_parameters1} \\
 		
 \end{table}%

 %

 
 \begin{table}[h!] \centering
 	\caption{The eigenvalues corresponding to the critical points of the system (\ref{autonomous1}) are presented, where $\Delta=1-3 \beta +3 \gamma$ }%
 	\begin{tabular}
 		[c]{cccccccc}\hline\hline
 		\textbf{Critical Point}&$\mathbf{\lambda_1}$& $\mathbf{\lambda_2}$ & $\mathbf{\lambda_3}$ & 
 		\\\hline
 		$P_1  $ & $2$ & $2$  & $3 \beta -3 \gamma +1$
 		\\ \\
 		$P_2  $ & $\frac{b-1}{b}$ & $-\frac{2 (b+1)}{b}$ & $3 \beta -\frac{2}{b}-3 \gamma -1$ \\ \\
 		$P_3  $ & $5-9 b^2$ & $2(1- b^2)$ & $1-4 b^2+3 \beta -3 \gamma $ \\ \\
 		$P_4  $ & $1-3 \beta +3 \gamma $ & $3\gamma-3 \beta -1$ & $-\frac{3}{2} (\beta -\gamma -1)$ \\ \\
 		$P_5  $ & $-4$ & $-2$ & $\frac{3}{2} (\beta -\gamma -1)$ \\ \\
 		$P_6  $ & $-\frac{\Delta}{8} \left(9 b^2 \Delta^2-20\right)$ & $\frac{1}{2} \left(-b^2 \Delta^3-6 \beta +6 \gamma -2\right)$ & $\frac{1}{4} \left(-b^2 \Delta^3-6 \beta +6 \gamma +6\right)$ \\ \\
 		$P_7  $ & $-4$ & $\frac{1}{4} \left[ -\sqrt{\left(b^2-1\right) \left\lbrace b^2 \Delta^2-(4+\Delta) ^2\right\rbrace }+b^2\Delta-\Delta\right] $ & $\frac{1}{4} \left[ \sqrt{\left(b^2-1\right) \left\lbrace b^2 \Delta^2-(4+\Delta)^2\right\rbrace }+b^2 \Delta-\Delta\right] $ 
 		\\
 		\\\hline\hline
 	\end{tabular}
 	\label{eigenvalues1} \\
 	
\end{table}%
%
\subsubsection{Local stability of critical points: Phase space analysis }\label{event phase space analysis}
In this section, we perform phase space analysis of the model and we present the local stability of the critical points by evaluating the eigenvalues of the linearized Jacobian matrix displayed in Table \ref{eigenvalues1}. Here, eigenvalues for all the critical points are non-zero, so the points are hyperbolic type. Therefore, linear stability theory is sufficient to achieve the stability of the critical points locally. The phase space structure of the three dimensional system can be found directly by observing the  associated eigenvalues of each critical point.

\begin{itemize}
\item The critical point $P_1$ exists for all free parameters ($\alpha$, $ \beta$, $\gamma$ and $b$) in the phase space $x-y-z$. The point represents completely radiation dominated solution in phase space ($\Omega_{r}=1,~\Omega_{m}=0,~\Omega_{d}=0$) (see table \ref{physical_parameters1}). Acceleration is not possible here ($q=1$). Stability of the point can be found by analysing the eigenvalues in Table \ref{eigenvalues1}. There exists two dimensional unstable sub-manifold in the eigendirections $\lambda_1=2$ and $\lambda_{2}=2$. Hence, the point will always be unstable in nature. 
We observe that the point $P_{1}$ will be unstable source (i.e. $\lambda_1>0$, $\lambda_{2}>0,\lambda_3 >0$) for $\gamma <\frac{1}{3} (3 \beta +1) $ and saddle (i.e. $\lambda_1>0$, $\lambda_{2}>0,\lambda_3 <0$) for $\gamma >\frac{1}{3} (3 \beta +1)$.
	
\item The critical point $P_2$ also exists for all free parameters: $\alpha$, $ \beta$, $\gamma$ and $b$ in the phase space $x-y-z$. It is completely HDE dominated solution as ($\Omega_{r}=0,~\Omega_{m}=0,~\Omega_{d}=1$ see table \ref{physical_parameters1}) and the HDE behaves as perfect fluid in nature $\omega_d=-\frac{b+2}{3b}$. Particular choice of $b$ signifies the nature of HDE for this critical point. HDE can mimic cosmological constant for $b=1$ and phantom for $b<1$. Acceleration exists for $b>0$. The point $P_{2}$ is stable sink (i.e. $\lambda_1<0$, $\lambda_{2}<0,\lambda_3 <	0$) for $\left(0<b<1~~\mbox{and}~~ \gamma >\frac{3 \beta  b-b-2}{3 b} \right) $ and is unstable source (i.e. $\lambda_1>0$, $\lambda_{2}>0,\lambda_3 >0$) for $\left(-1<b<0~~\mbox{and}~~ \gamma <\frac{3 \beta  b-b-2}{3 b}\right).$ Further, the point shows saddle like nature in the phase space for $\left( 0<b<1~~\mbox{and}~~\gamma <\frac{3 \beta  b-b-2}{3 b}\right).$ From the stability analysis, one may conclude that the point $P_2$ can show the late time acceleration when it  evolves in phantom era (i.e. $\lambda_1<0,  \lambda_2<0, \lambda_3<0~\mbox{and}~ \omega_{eff}<-1$) for
	$\left( 0<b<1~~\mbox{and}~~ \gamma >\frac{3 \beta  b-b-2}{3 b}\right) $.
Also, the point will tend to a late time de Sitter solution when $b\longrightarrow 1$ which corresponds to the acceleration evolution of the universe ($\Omega_{d}=1,~\omega_{eff}\approx-1,~q\approx-1$) where HDE evolves with equation of state $\omega_{d}\approx-1$.	
	
\item The critical point $P_3$ is a solution of combination of both the radiation and the HDE and it exist for free parameters $\alpha$, $ \beta$, $\gamma$ and $b$ with $0\leq b^{2}\leq 1$ in the phase space $x-y-z$. The HDE associated to the critical point behaves as cosmological constant ($\omega_{d}=-1$). 
	
 Acceleration will be achieved when $b^2>\frac{1}{2}$.
Eigenvalues of the linearised Jacobian matrix corresponding to the point will characterise the stability of the point. We find that the point $P_{3}$ will be unstable source (i.e. $\lambda_1>0$, $\lambda_{2}>0,\lambda_3 >0$) for $\left\lbrace -\frac{\sqrt{5}}{3}<b<\frac{\sqrt{5}}{3}~~\mbox{and}~~ \gamma <\frac{1}{3} \left(-4 b^2+3 \beta +1\right)\right\rbrace $. The point $P_3$ represents decelerated unstable source (i.e. $\lambda_1>0,  \lambda_2>0, \lambda_3>0~\mbox{and}~ \omega_{eff}>-\frac{1}{3}$) for
	$\left\lbrace -\frac{1}{\sqrt{2}}<b<\frac{1}{\sqrt{2}}~~\mbox{and}~~ \gamma <\frac{1}{3} \left(-4 b^2+3 \beta +1\right)\right\rbrace  $.
Otherwise, the point will be saddle like solution in the phase space. However, the point can never be stable solution. The point will come into DE domination for $b\longrightarrow 1$ ($\Omega_{r}\approx0,~\Omega_{m}=0,~\Omega_{d}\approx1$ see table \ref{physical_parameters1}). Unfortunately, it cannot show the late time solution. Therefore, it is not physically interested for this case. On the other hand, when $b\longrightarrow 0$, the point will become radiation dominated ($\Omega_{r}\approx 1$) and for this case, the point will show decelerated source like solution.	
\begin{figure}
	\centering
	\subfigure[]{%
		\includegraphics[width=8.5cm,height=10.0cm]{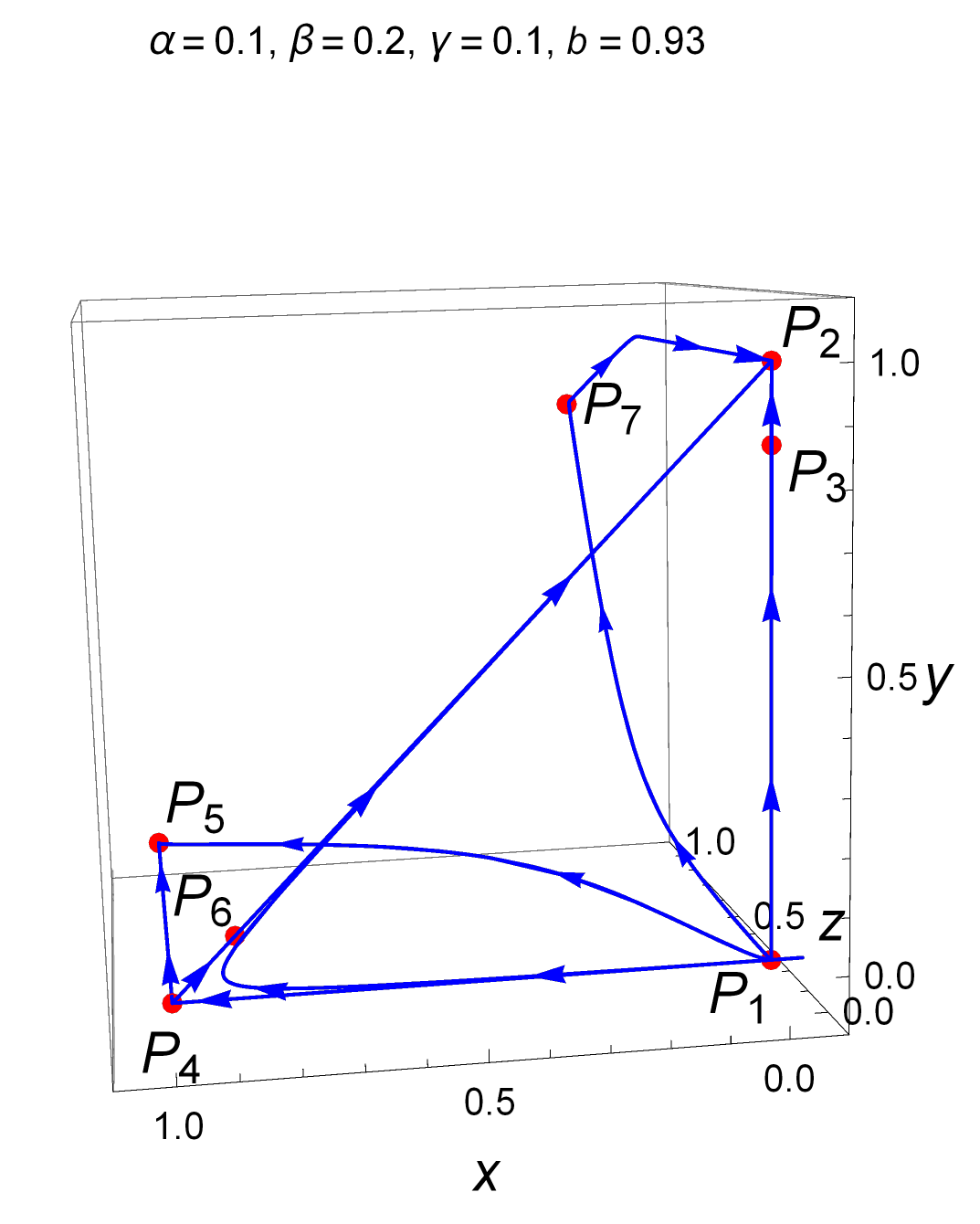}\label{Event_3D}}
	\qquad
	\subfigure[]{%
		\includegraphics[width=8.5cm,height=8.5cm]{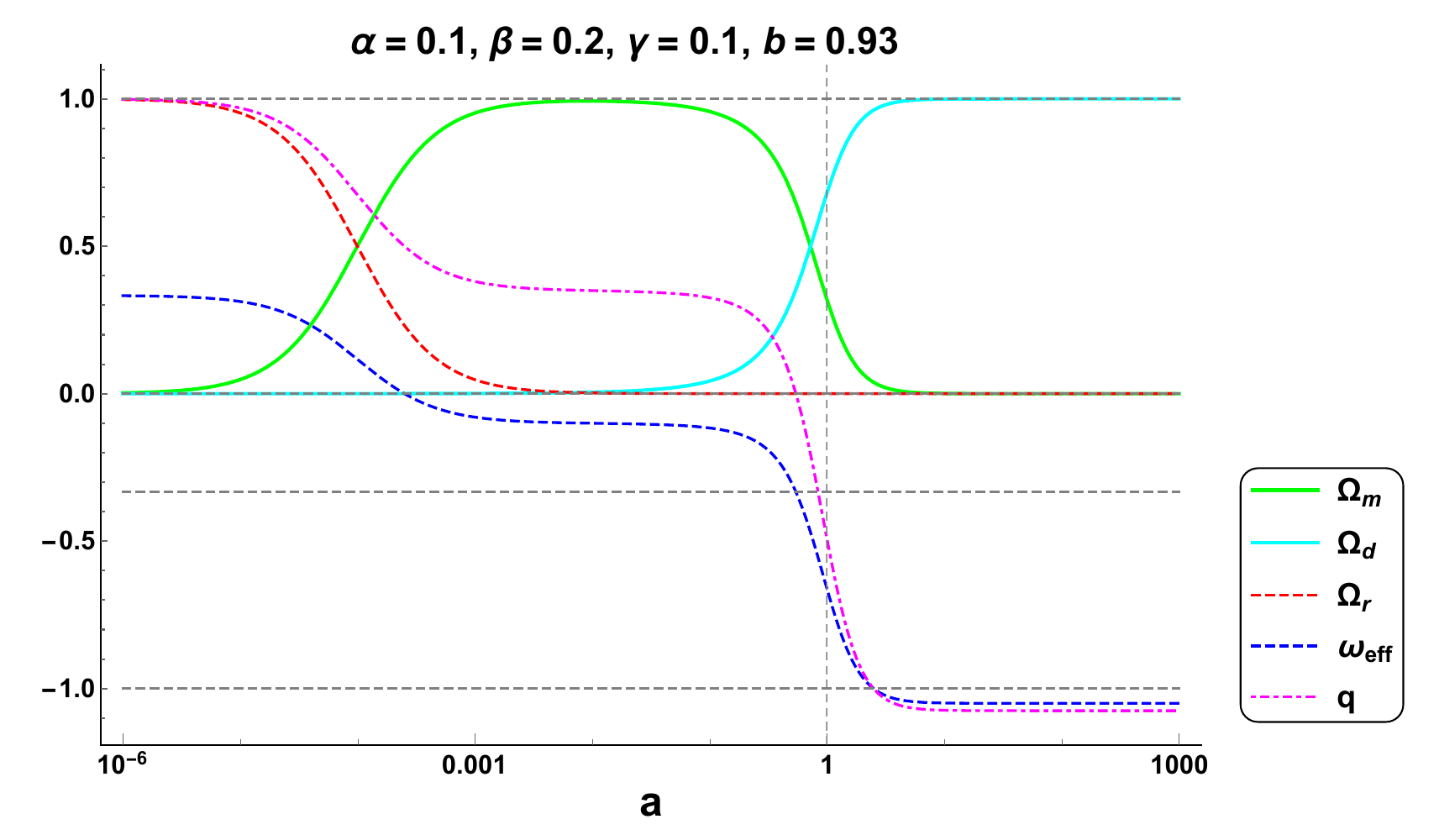}\label{Event_evolution}}
	\caption{The trajectories of 3D autonomous system (\ref{autonomous1}) and the time- evolution of the cosmological parameters are plotted for parameter values $\alpha=0.1,~~\beta=0.2,~~\gamma=0.1,~~b=0.93$. Sub-figure \ref{Event_3D} exhibits that the point $P_1$ is a radiation dominated past attractor, $P_2$ is DE dominated late time attractor. But, the point $P_3$ is DE dominated accelerated saddle-like solution and $P_4$ is DM dominated decelerated saddle-like solution, $P_5$ is DM dominated accelerated stable, $P_6$ is DM dominated decelerated saddle-like solution and $P_7$ represents the DE dominated accelerated saddle-like solution. Sub-figure \ref{Event_evolution} shows the evolution of cosmological parameters where trajectories are attracted in DE dominated phantom era. }
	\label{P2deSitter}
\end{figure} 
	
\item The dark matter dominated ($\Omega_{m}=1$) solution represented by the critical point $P_4$ will always exist for all $\alpha$, $ \beta$, $\gamma$ and $b$ in the phase space $x-y-z$. Acceleration will be achieved for $\gamma<\beta-\frac{1}{3}$. From the eigenvalues in Table \ref{eigenvalues1}, the point $P_{4}$ will show an unstable source (i.e. $\lambda_1>0$, $\lambda_{2}>0,\lambda_3 >0$) for $\left(\gamma >\frac{1}{3} (3 \beta +1)\right)$. 
The point $P_4$ represents decelerated unstable source (i.e. $\lambda_1>0,  \lambda_2>0, \lambda_3>0~\mbox{and}~ \omega_{eff}>-\frac{1}{3}$) for
	$\left( \gamma >\frac{1}{3} (3 \beta +1)\right) $.
Note that for $\beta=\gamma$, the critical point will mimic the dust dominated ($\omega_{eff}=0$) decelerated evolution ($q=\frac{1}{2}$)	which describes the transient phase of the universe (for this case, the point is saddle like solution).
	
\item The critical point $P_5$ also a DM dominated solution ($\Omega_{m}=1$) which will exist for  $(\alpha -\beta +\gamma +1\neq 0)$ and for all $b$.
The point will be a saddle solution (i.e. $\lambda_1<0$, $\lambda_{2}<0,\lambda_3 >0$) for :\\
	$
	(i)~~(\alpha \leq 0~~\mbox{and}~~ \gamma <\beta -1),~~\mbox{or}~~\\
	(ii)~~ \alpha >0~~\mbox{and}~~ (\gamma <-\alpha +\beta -1~~\mbox{or}~~ -\alpha +\beta -1<\gamma <\beta -1)$\\
The universe is always accelerated near the universe (since $q=-1,~\omega_{eff}=-1$). The solution is unphysical because it represents matter dominated accelerated intermediate phase which does not fit to the observations.
\item HDE- DM scaling solution ($\Omega_{d}\sim \Omega_{m}$) represented by the critical point $P_6$ exists for all $\alpha$, and  $\Big[b=0~~\mbox{or}~~ \left(\frac{2}{\left| b\right| }+3 \beta \geq 3 \gamma +1~~\mbox{and}~~  2 \sqrt{\frac{1}{b^2}}+3 \gamma +1\geq 3 \beta \right)  \Big]$ in the phase space $x-y-z$. The ratio of DE and DM is $r=\frac{\Omega_{d}}{\Omega_{m}}=\frac{b^2 \Delta^2}{4-b^2\Delta^2}$. The  point will become completely DE dominated for $\Big[ \left(b=-\frac{2}{-3 \beta +3 \gamma +1}~~\mbox{or}~~ \frac{2}{-3 \beta +3 \gamma +1}=b\right)$ \\$~\mbox{and}~~\beta \neq \gamma +\frac{1}{3}  \Big]$ ($\Omega_{r}=0,~\Omega_{m}=0,~\Omega_{d}=1$ see table \ref{physical_parameters1}). The HDE corresponds to an exotic type fluid which evolves with equation of state $\omega_{d}=\frac{\gamma  \left\lbrace 4-b^2 (-3 \beta +3 \gamma +1)^2\right\rbrace }{b^2 (-3 \beta +3 \gamma +1)^2}+\beta -\gamma -\frac{2}{3}$. Also, physical parameter $q=\frac{1}{4} (3 \beta -3 \gamma -1) \left\lbrace b^2 (-3 \beta +3 \gamma +1)^2-2\right\rbrace $ suggests that acceleration of the universe near the point is possible depending on model parameters $\beta$, $\gamma$ and $b$.\\
	
Eigenvalues of the Jacobian matrix signify it to be a stable solution (i.e. $\lambda_1<0$, $\lambda_{2}<0,\lambda_3 <0$) for the following conditions :
	
	\begin{enumerate}
		\item  $2 \sqrt{\frac{1}{(-3 \beta +3 \gamma +1)^2}}+b\geq 0 :~~\\
		(i)~~ \left(\sqrt{6} \sqrt{\frac{\beta -\gamma -1}{(3 \beta -3 \gamma -1)^3}}+b<0~~\mbox{and}~~ \beta +\frac{1}{3}<\gamma ~~\mbox{and}~~ \gamma <\beta +\frac{17}{3}\right),~~\mbox{or}~~\\ (ii)~~\left(\frac{2 \sqrt{5}}{-3 \beta +3 \gamma +1}+3 b<0~~\mbox{and}~~ \beta +\frac{17}{3}\leq \gamma \right)$
		\item  $\left\lbrace b<\sqrt{6} \sqrt{\frac{\beta -\gamma -1}{(3 \beta -3 \gamma -1)^3}}~~\mbox{and}~~ \sqrt{6} \sqrt{\frac{\beta -\gamma -1}{(3 \beta -3 \gamma -1)^3}}+b>0~~\mbox{and}~~ \beta >\gamma +1\right\rbrace $ 
		\item  $b\leq 2 \sqrt{\frac{1}{(-3 \beta +3 \gamma +1)^2}} :~~\\
		(i)~~ \left(\beta +\frac{1}{3}<\gamma ~~\mbox{and}~~ \sqrt{6} \sqrt{\frac{\beta -\gamma -1}{(3 \beta -3 \gamma -1)^3}}<b~~\mbox{and}~~\gamma <\beta +\frac{17}{3}\right),~~\mbox{or}~~\\
		(ii)~~ \left(\frac{2 \sqrt{5}}{-3 \beta +3 \gamma +1}<3 b~~\mbox{and}~~ \beta +\frac{17}{3}\leq \gamma \right)$
	\end{enumerate}
and will be unstable source (i.e. $\lambda_1>0$,   $\lambda_{2}>0,\lambda_3 >0$) for the following conditions :
	
	\begin{enumerate}
		\item  $2 \sqrt{\frac{1}{(-3 \beta +3 \gamma +1)^2}}+b\geq 0 :~~\\
		(i)~~ \left(\sqrt{\frac{6 \beta -6 \gamma +2}{(3 \beta -3 \gamma -1)^3}}+b<0~~\mbox{and}~~\gamma +1<\beta ~~\mbox{and}~~ \beta \leq \gamma +\frac{19}{3}\right),~~\mbox{or}~~\\
		(ii)~~ \left(\beta >\gamma +\frac{19}{3} ~~\mbox{and}~~ 3 b \beta +\frac{2 \sqrt{5}}{3}<3 b \gamma +b\right)$
		\item  $ \left(b<\sqrt{\frac{6 \beta -6 \gamma +2}{(3 \beta -3 \gamma -1)^3}}~~\mbox{and}~~ \beta +\frac{1}{3}<\gamma ~~\mbox{and}~~ \sqrt{\frac{6 \beta -6 \gamma +2}{(3 \beta -3 \gamma -1)^3}}+b>0\right)$
		\item  $ b\leq 2 \sqrt{\frac{1}{(-3 \beta +3 \gamma +1)^2}} :~~\\
		(i)~~\left(\sqrt{\frac{6 \beta -6 \gamma +2}{(3 \beta -3 \gamma -1)^3}}<b~~\mbox{and}~~ \gamma +1<\beta ~~\mbox{and}~~\beta \leq \gamma +\frac{19}{3}\right),~~\mbox{or}~~\\
		(ii)~~ \left(\beta >\gamma +\frac{19}{3}~~\mbox{and}~~ b (9 \beta -9 \gamma -3)>2 \sqrt{5}\right)$.
	\end{enumerate}
Finally, the point $P_6$ will be a saddle like solution (i.e. $\lambda_1<0$, $\lambda_{2}>0,\lambda_3 >0$) for the following conditions
	$\left\lbrace \beta >\gamma +\frac{19}{3}~~\mbox{and}~~ \left(\frac{2 \sqrt{5}}{-9 \beta +9 \gamma +3}<b<-\frac{\sqrt{6 \beta -6 \gamma +2}}{(3 \beta -3 \gamma -1)^{3/2}}~~\mbox{or}~~ \frac{\sqrt{6 \beta -6 \gamma +2}}{(3 \beta -3 \gamma -1)^{3/2}}<b<-\frac{2 \sqrt{5}}{-9 \beta +9 \gamma +3}\right) \right\rbrace $.
There will be acceleration of the evolution (i.e. $\omega_{eff}<-\frac{1}{3}$) near the point for the following conditions :	\begin{enumerate}
		\item  $\left\lbrace \beta <\gamma +\frac{1}{3}~~\mbox{and}~~ \left(\sqrt{2}<b (3 \beta -3 \gamma -1)\leq 2~~\mbox{or}~~ -2\leq b (3 \beta -3 \gamma -1)<-\sqrt{2}\right)\right\rbrace,~~\mbox{or}~~ $
		\item  $ \frac{\sqrt{2}}{-3 \beta +3 \gamma +1}<b<\frac{\sqrt{2}}{3 \beta -3 \gamma -1}$
	\end{enumerate}
From the above analysis, we can conclude that the point represents the late time accelerated evolution of the universe attracted only in phantom era (i.e. $\lambda_1<0$, $\lambda_2<0, \lambda_3<0~\mbox{and}~ \omega_{eff}<-1$) for  ~
	\begin{enumerate}
		\item  $2 \sqrt{\frac{1}{(-3 \beta +3 \gamma +1)^2}}+b\geq 0 :~~\\
		(i)~~ \left(\sqrt{6} \sqrt{\frac{\beta -\gamma -1}{(3 \beta -3 \gamma -1)^3}}+b<0~~\mbox{and}~~ \beta +\frac{1}{3}<\gamma ~~\mbox{and}~~ \gamma <\beta +\frac{17}{3}\right),~~\mbox{or}~~\\
		(ii)~~ \left(\frac{2 \sqrt{5}}{-3 \beta +3 \gamma +1}+3 b<0~~\mbox{and}~~ \beta +\frac{17}{3}\leq \gamma \right)$
		\item $ \left(b<\sqrt{6} \sqrt{\frac{\beta -\gamma -1}{(3 \beta -3 \gamma -1)^3}}~~\mbox{and}~~ \sqrt{6} \sqrt{\frac{\beta -\gamma -1}{(3 \beta -3 \gamma -1)^3}}+b>0~~\mbox{and}~~ \beta >\gamma +1\right)$
		\item  $b\leq 2 \sqrt{\frac{1}{(-3 \beta +3 \gamma +1)^2}} :~~\\
		(i)~~ \left(\beta +\frac{1}{3}<\gamma ~~\mbox{and}~~ \sqrt{6} \sqrt{\frac{\beta -\gamma -1}{(3 \beta -3 \gamma -1)^3}}<b~~\mbox{and}~~ \gamma <\beta +\frac{17}{3}\right),~~\mbox{or}~~\\
		(ii)~~ \left(\frac{2 \sqrt{5}}{-3 \beta +3 \gamma +1}<3 b~~\mbox{and}~~ \beta +\frac{17}{3}\leq \gamma \right)$
	\end{enumerate}
The point will be decelerated unstable source (i.e. $\lambda_1>0$,  $\lambda_2>0, \lambda_3>0~\mbox{and}~ \omega_{eff}>-\frac{1}{3}$) for ~
	\begin{enumerate}
		\item  $2 \sqrt{\frac{1}{(-3 \beta +3 \gamma +1)^2}}+b\geq 0 :~~\\
		(i)~~ \left(\sqrt{\frac{6 \beta -6 \gamma +2}{(3 \beta -3 \gamma -1)^3}}+b<0~~\mbox{and}~~ \gamma +1<\beta ~~\mbox{and}~~ \beta \leq \gamma +\frac{19}{3}\right),~~\mbox{or}~~\\
		(ii)~~ \left(\beta >\gamma +\frac{19}{3}~~\mbox{and}~~ 3 b \beta +\frac{2 \sqrt{5}}{3}<3 b \gamma +b\right)$
		\item  $ \left(b<\sqrt{\frac{6 \beta -6 \gamma +2}{(3 \beta -3 \gamma -1)^3}}~~\mbox{and}~~ \beta +\frac{1}{3}<\gamma ~~\mbox{and}~~ \sqrt{\frac{6 \beta -6 \gamma +2}{(3 \beta -3 \gamma -1)^3}}+b>0\right)$
		\item $ b\leq 2 \sqrt{\frac{1}{(-3 \beta +3 \gamma +1)^2}} :~~\\
		(i)~~ \left(\sqrt{\frac{6 \beta -6 \gamma +2}{(3 \beta -3 \gamma -1)^3}}<b~~\mbox{and}~~ \gamma +1<\beta ~~\mbox{and}~~ \beta \leq \gamma +\frac{19}{3}\right),~~\mbox{or}~~\\
		(ii)~~ \left(\beta >\gamma +\frac{19}{3}~~\mbox{and}~~ b (9 \beta -9 \gamma -3)>2 \sqrt{5}\right)$
	\end{enumerate}
	The point will be decelerated saddle (i.e. $\lambda_1<0$,  $\lambda_2>0, \lambda_3>0~\mbox{and}~ \omega_{eff}>-\frac{1}{3}$) for ~\\
	$\left\lbrace \beta >\gamma +\frac{19}{3}~~\mbox{and}~~ \left(\frac{2 \sqrt{5}}{-9 \beta +9 \gamma +3}<b<-\frac{\sqrt{6 \beta -6 \gamma +2}}{(3 \beta -3 \gamma -1)^{3/2}}~~\mbox{or}~~ \frac{\sqrt{6 \beta -6 \gamma +2}}{(3 \beta -3 \gamma -1)^{3/2}}<b<-\frac{2 \sqrt{5}}{-9 \beta +9 \gamma +3}\right) \right\rbrace $.
\item Another scaling solution ($\Omega_{d}\sim \Omega_{m}$) described by the critical point $P_7$ will exist for $\Big[-1\leq b\leq 1~~\mbox{and}~~ (\gamma <-\alpha +\beta -1~~\mbox{or}~~ \gamma >-\alpha +\beta -1)\Big]$ in the phase space $x-y-z$. The ratio of DE and DM is $r=\frac{\Omega_{d}}{\Omega_{m}}=\frac{b^2}{1-b^2}$.
The point will become completely DE dominated for $b=1$ ($\Omega_{r}=0$,~$\Omega_{m}=0,~\Omega_{d}=1$ see table \ref{physical_parameters1}). The DE behaves as exotic type of fluid as $\omega_{d}=\left(\frac{1}{b^2}-1\right) \gamma -1$. The point will always be accelerating $q=-1$ and evolving in cosmological constant era.
The point $P_{7}$ will be saddle (i.e. $\lambda_1<0$, $\lambda_{2}<0,\lambda_3 >0$) for \\
		$-1<b<1~~:\\
		(i)~~\left\lbrace  \alpha <0~~\mbox{and}~~ (\beta -1<\gamma <-\alpha +\beta -1~~\mbox{or}~~ \gamma >-\alpha +\beta -1)\right\rbrace ,~~\mbox{or}~~\\
		(ii)~~ (\alpha \geq 0~~\mbox{and}~~ \gamma >\beta -1)$.
Note that the point can never be stable solution in the phase space so, it is not a physically relevant solution at late times.
 

It is worthy to note that the parameters constraints $\beta=\gamma$ gives some insightful results of the model where the restrictions fully depending on $b$. The point $P_1$ represents the radiation dominated, decelerated evolution ($\omega_{eff}=\frac{1}{3}$, $q=1$) at early times (because the point is source in phase space).\\

$P_2$ corresponds to HDE dominated ($\Omega_{d}=1$), accelerated late-time evolution in phantom era for positive $b$. In particular, for $b\longrightarrow 1$ the point describes late-time de Sitter evolution of the universe ($\Omega_{d}=1, $$\omega_{eff} \approx -1$, $q\approx-1$), where HDE behaves as cosmological constant ($\omega_{d}=-1$).\\

Critical point $P_3$ will show radiation dominated solution for $b\longrightarrow 0$ and it represents early (it becomes source in phase space) decelerated evolution of the universe evolving in radiation era ($\omega_{eff}=\frac{1}{3}$) as is shown by point $P_1$. On the other hand, for $b\longrightarrow 1$ the point $P_3$ will represent late-time accelerated de Sitter solution ($\omega_{eff}\approx -1,~ q\approx -1$).\\

Dark matter dominated ($\Omega_{m}=1$) point $P_4$ evolves in dust era ($\omega_{eff}=0$) and it has always decelerating nature $q=\frac{1}{2}$ and it represents (being a saddle solution in phase space) a transient phase of the universe.\\

The point $P_5$ is also a matter dominated point ($\Omega_{m}=1$) and it shows a late-time (all the eigenvalues have negative real parts) accelerated evolution ($\omega_{eff}=-1,~ q=-1$). This result does not fit to the observational data.\\

The scaling solution represented by $P_6$ will always be a saddle like solution in phase space. In particular, when $b\longrightarrow 1$, the point will show a DM dominated ($\Omega_{m}\approx \frac{3}{4}$, $\Omega_{d}\approx \frac{1}{4}$) decelerated evolution and it represent the intermediate phase of the universe (because the point is saddle like solution in phase space). On the other hand, for $b\longrightarrow 0$, the point will become DM dominated ($\Omega_{m}\approx 1,~ \Omega_{d}\approx 0$) solution evolving in dust era ($\omega_{eff}\approx0,~q\approx \frac{1}{2}$) corresponding to a decelerated transient phase of the universe (because the point is saddle in phase space). This nature of the critical point $P_6$ is shown in fig.(\ref{Event_3D1}) for $\alpha=0.1,~~\beta=0.1,~~\gamma=0.1,~~b=0.93$. \\

Another scaling solution $P_7$ shows an unstable solution in phase space. However, for $b\longrightarrow 1$, the point will become HDE dominated solution and it will have one non-empty stable sub manifold corresponding to negative eigenvalue ($\lambda_1=-4,~\lambda_2 \approx 0, \lambda_3\approx 0$). Then the point can be a de Sitter solution in phase space which is accelerating $\Omega_{d}\approx1,~ \Omega_{m}\approx 0, \omega_{eff} =-1,~q=-1$ where the HDE behaves as cosmological constant $\omega_{d}\approx -1$. However, for $b\longrightarrow0$ the point will become DM dominated saddle solution, but always accelerating. So, it is not physically  interested for this case.
A 3D phase plot is drawn in fig. (\ref{Event_3D1}) for $\alpha=0.1,~~\beta=0.1,~~\gamma=0.1,~~b=0.93$ to make out the dynamics of the model where we observe all the critical points.

\begin{figure}
	\centering
     \includegraphics[width=12cm,height=12cm]{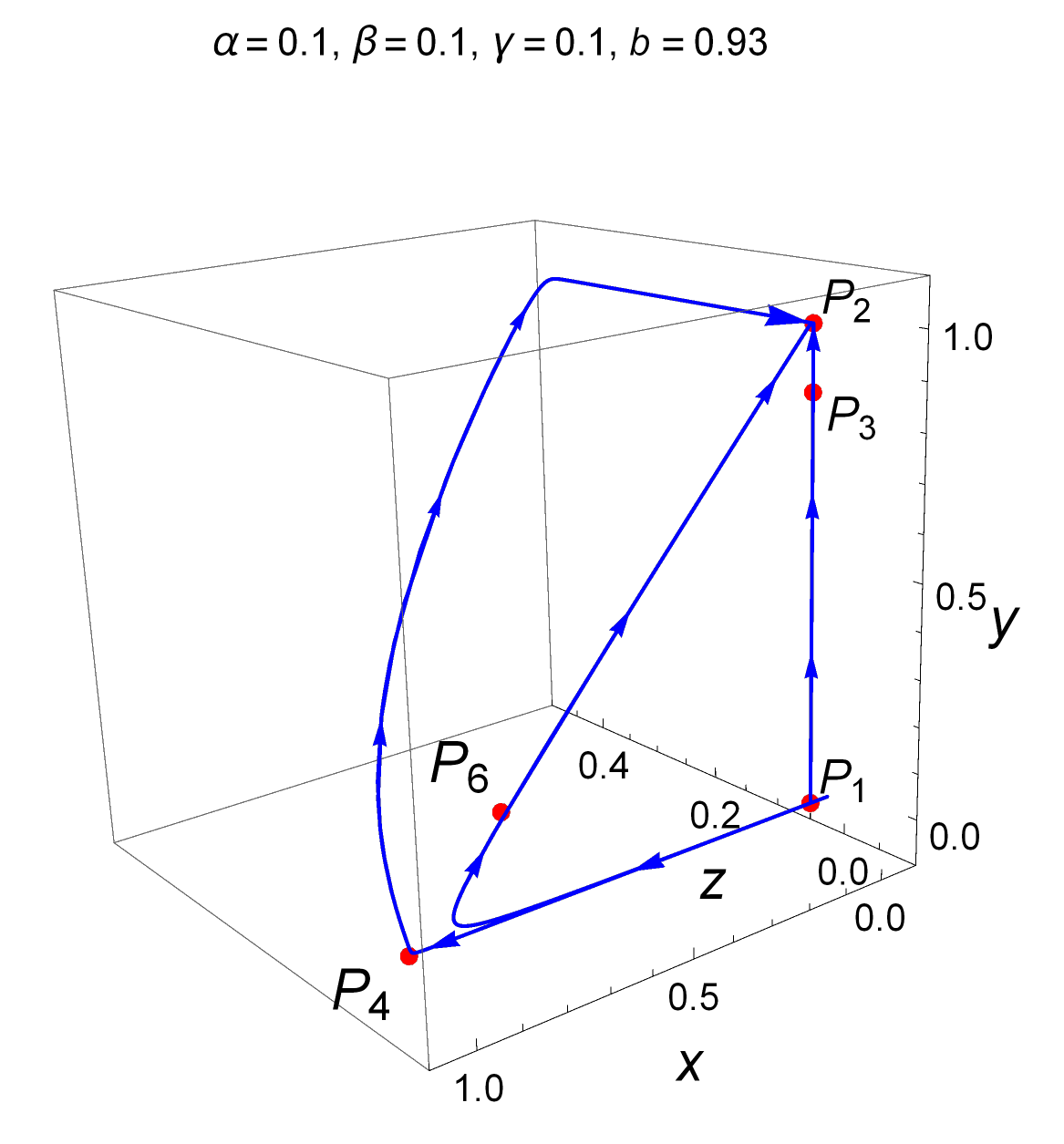}
	\caption{The figures shows that trajectories of the autonomous system (\ref{autonomous1}) for parameter values $\alpha=0.1,~~\beta=0.1,~~\gamma=0.1,~~b=0.93$. It shows that $P_1$ represents radiation dominated source, $P_2$ corresponds to DE dominated accelerated late time attractor, $P_3$ is DE dominated accelerated saddle-like solution, $P_4$ is DM dominated decelerated saddle-like solution and $P_6$ is DM dominated decelerated saddle-like solution.}
	\label{Event_3D1}
\end{figure} 
\begin{figure}
	\centering
	\includegraphics[width=8.5cm,height=8.5cm]{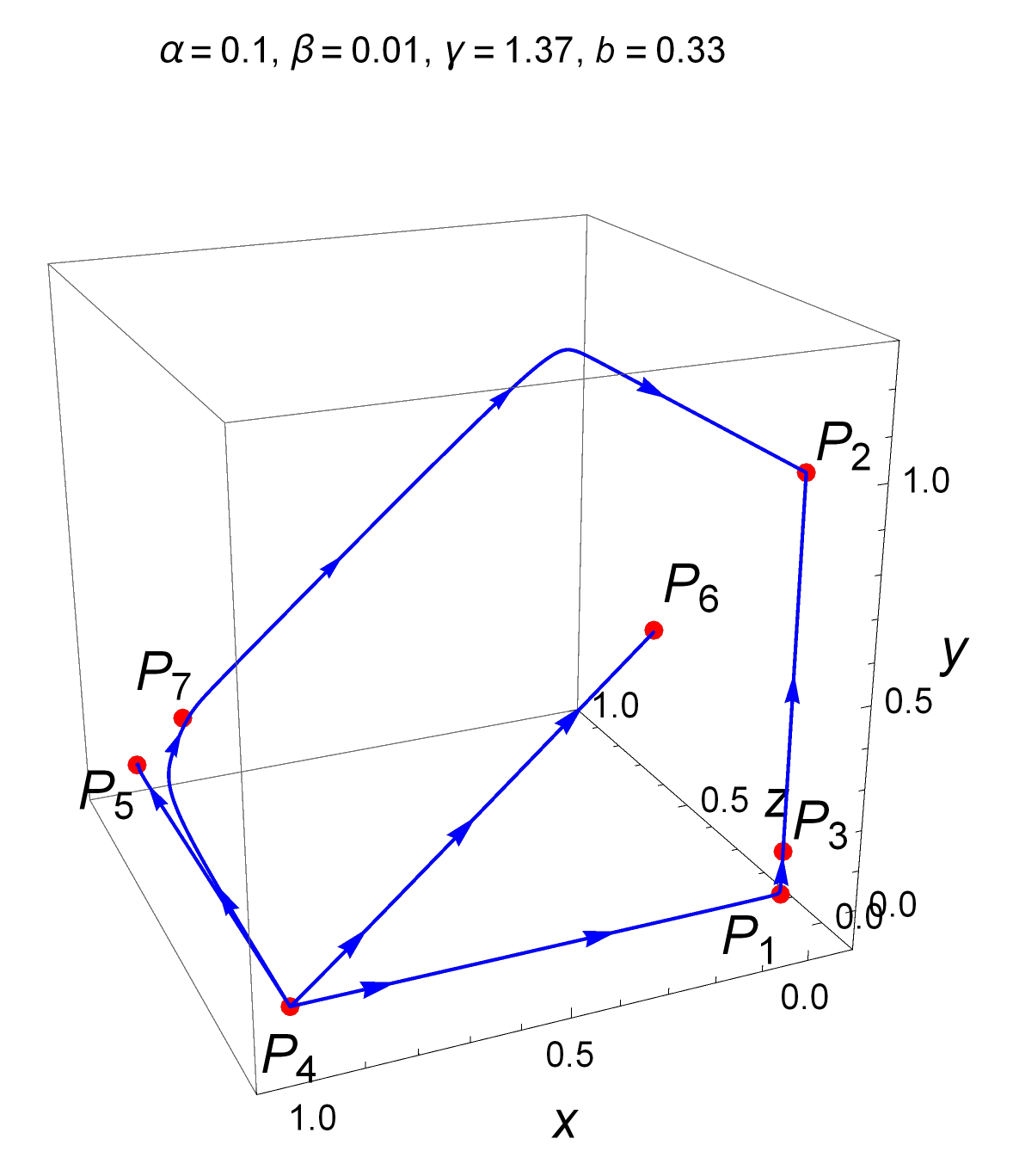}
	\caption{The figure shows the phase space trajectories in 3D for the autonomous system (\ref{autonomous1}) for $\alpha=0.1,~~\beta=0.01,~~\gamma=1.37,~~b=0.33$. It shows different nature of critical points as follows: the point $P_1$ represents radiation dominated saddle-like solution, $P_2$ corresponds to DE dominated accelerated late time attractor, $P_3$ is radiation dominated decelerated saddle-like solution, $P_4$ is DM dominated decelerated source solution, $P_5$ is DM dominated accelerated stable, and the point $P_6$ represents the scaling solution (DE dominated) corresponding to late-time accelerated attractor with $\Omega_{m}=0.297,~~\Omega_{d}=0.703,~~\Omega_{r}=0,~~\omega_{d}=-1.446,~~\omega_{eff}=-1.019,~~q=-1.029$. The point $P_7$ represents DM dominated accelerated saddle-like solution.}
	\label{P6_3D}
\end{figure} 

\end{itemize}
\begin{figure}
	\centering
	\includegraphics[width=16cm,height=9cm]{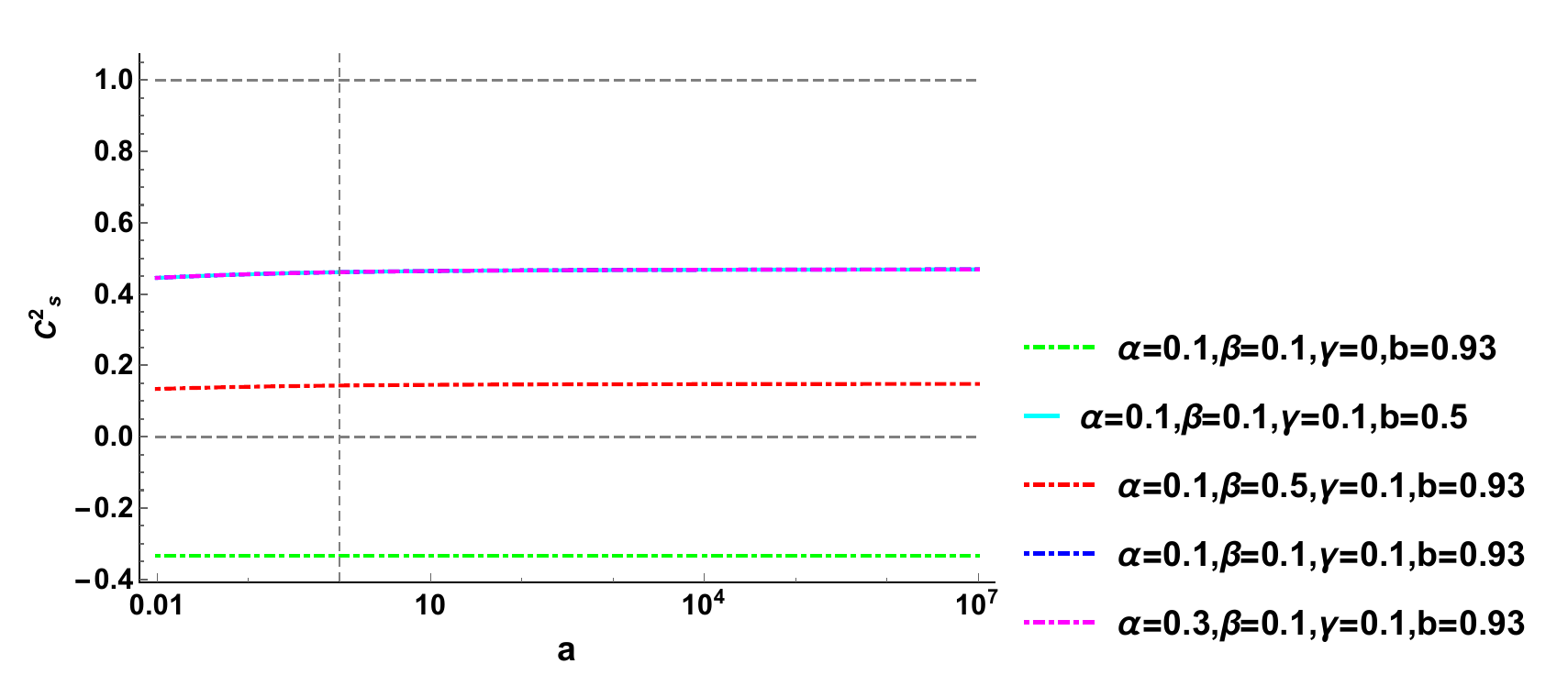}
	
	\caption{The figure shows sound speed curves of HDE with future event horizon as IR cutoff for different choices of model parameters.
	}
	\label{event_evolution}
\end{figure}  

\subsubsection{Classical stability of the model}

We now investigate the stability of the model by finding the squared sound speed for the model of HDE with Event horizon as IR cutoff. By using the Eqns. (\ref{Event omega}), (\ref{variables1}), (\ref{evolution Event}), we obtain sound speed from the expression (\ref{sound speed formula}) in terms of dynamical variables $x(N)$, y(N) and z(N) as follows:
\begin{align*}
	C_{s}^{2}=\Big[ b^2 \left\lbrace 9 \gamma  x (\beta -\gamma +z (\alpha -\beta +\gamma +1)-1)-2 y  (z-1)\right\rbrace +b y^{3/2} \left\lbrace x (3 \beta -3 \gamma +z (3 \alpha -3 \beta +3 \gamma -1)+1)-2 y(z-1)\right\rbrace  \\ -2 (y -3) y^2 (z-1) \Big] \div \left\lbrace 6 by  (z-1) \left(b-\sqrt{y }\right)\right\rbrace 
\end{align*}	  
Fig. (\ref{event_evolution}) indicates the evolution of squared speed of sound for different values of parameters. It shows that for non-interacting case i.e., taking $\gamma=0$, the model is unstable while for non-zero coupling term ($\gamma\neq 0$) the model is stable throughout the evolution of the universe.


\subsection{Global Dynamics of the Model }\label{global}

In this section, we perform the global stability analysis of the system (\ref{autonomous1}) around the
critical point $P_2$. The global stability of the critical point means  the
system (\ref{autonomous1}) with arbitrary positive initial values  all the solution trajectories converge to a unique critical point.

\begin{thm}
	Suppose that
	
	$2\left(1+\frac{1}{b}\right)+3\gamma>2\epsilon_2\left(1+\frac{\sqrt{\epsilon_2}}{b}\right)+\epsilon_1\left[1+3\beta+\frac{3\alpha \epsilon_3}{1-\epsilon_3})\right]$, where $\epsilon_{i}$'s are sufficiently small positive real numbers and prescribed in the proof of theorem.
	Then the  critical point $P_2(0, 1, 0)$ of the system (\ref{autonomous1}) implies its globally asymptotically stable around the critical point $P_2$.
	\label{Theorem:7}
\end{thm}

\textbf{Proof.}
We consider the following  positive definite Lyapunov function

$$L=\frac{1}{2}(1-y)^2.$$

We compute the rate of change of $L$ along the solutions of (\ref{autonomous1}),  we get
$\frac{dL}{dN}=-(1-y)\frac{dy}{dN}$\\
$=-y(1-y) \Bigl[ 2(1-y)(1+\frac{\sqrt{y}}{b})-x(1+3\beta-3\gamma+\frac{3\alpha z}{1-z})\Bigr]. $\\
Now $\frac{dL}{dN}<0$, if $2(1-y)(1+\frac{\sqrt{y}}{b})-x(1+3\beta-3\gamma+\frac{3\alpha z}{1-z})>0$ since
$0<y<1$.
\\
After some algebraic calculations and using the inequalities $\epsilon_1<x<1$, $\epsilon_2<y<1$, $\epsilon_3<z<1$, with $\epsilon_{i}$'s $(i=1,2,3)$ are sufficiently smalls, we obtain

\begin{figure}[ht!]
	\centering
	\includegraphics[height=8.5cm , width=12.cm]{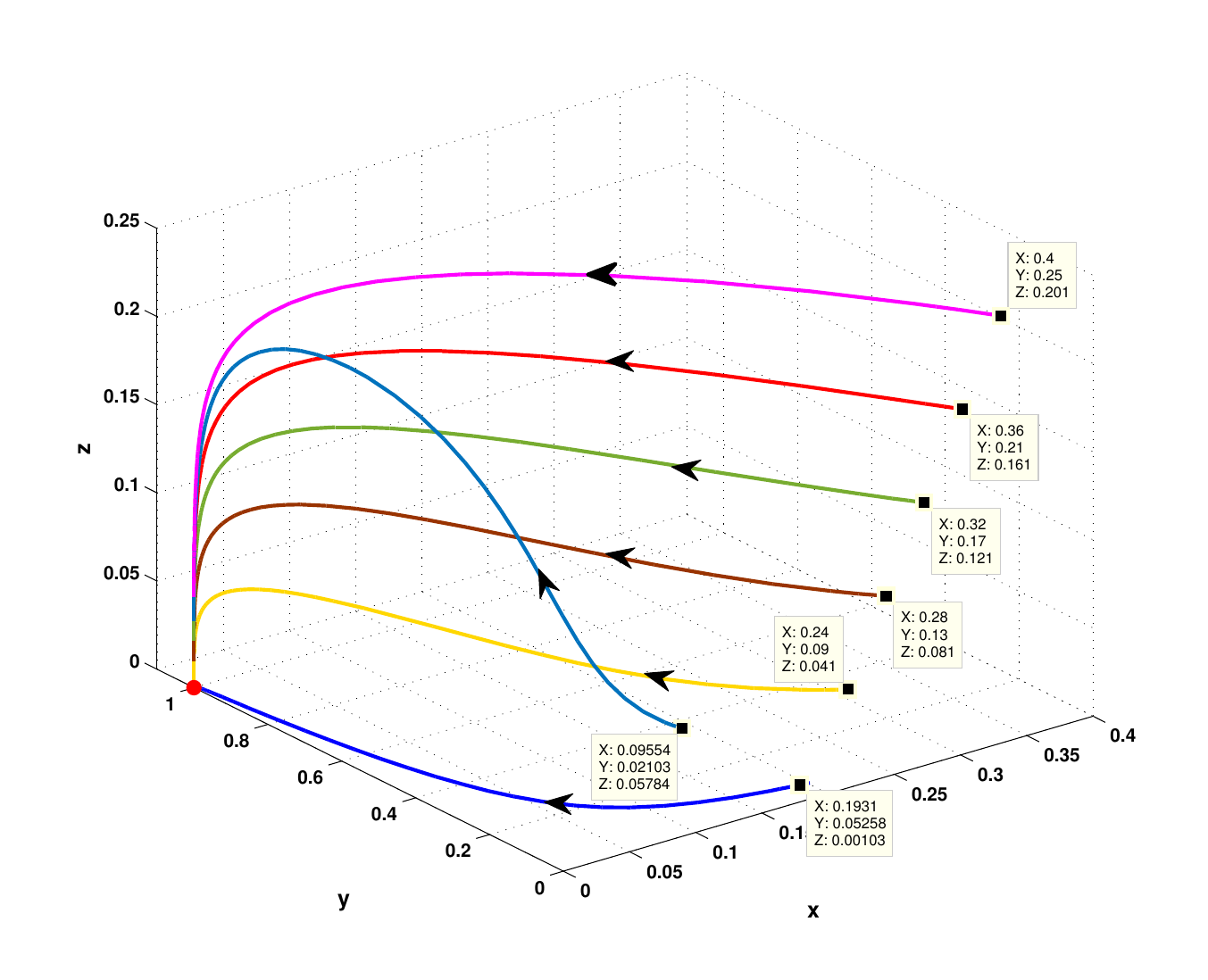}
	\caption{Figure shows the global stability of the critical point $P_2$  of the system (\ref{autonomous1}) for  $b=0.33$, $\alpha=0.1$, $\beta=0.01$, $\gamma=1.37$.
 Different colors are used to denote the trajectories starting from different initial values.
}
	\label{global stable}
\end{figure}

$\frac{dL}{dN}<0$ when $2\left(1+\frac{1}{b}\right)+3\gamma>2\epsilon_2\left(1+\frac{\sqrt{\epsilon_2}}{b}\right)+\epsilon_1\left[1+3\beta+\frac{3\alpha \epsilon_3}{1-\epsilon_3})\right]$.
\\
Thus if the condition of this theorem holds then the system (\ref{autonomous1}) is globally asymptotically stable around the critical point $P_2$. The fig. (\ref{global stable}) with the values of free parameters: $b=0.33$, $\alpha=0.1$, $\beta=0.01$, $\gamma=1.37$ shows that the point $P_2$ is a global attractor.

\section{Physical implications of critical points}\label{cosmological implications}
We have studied interacting HDE with creation rate $\Gamma=3\alpha H_0+3\beta H$ in model 2 with the help of dynamical analysis for Hubble horizon and the future event horizon as IR cutoffs separately in two subsections.
We have obtained the autonomous systems (\ref{autonomous}) and (\ref{autonomous1}) for two cutoffs respectively. We shall now discuss the cosmological implications of the critical points obtained from the systems. First system extracts two non-isolated set of critical points namely, $C_1$ and $C_2$ among which first one is normally hyperbolic set and from the stability analysis, we obtained a late-time attractor ($C_1$) corresponding to accelerating universe evolving different epochs such as in quintessence epoch, in cosmological constant era, or in phantom era. This scenario is shown in fig.(\ref{Point C1 C2}) for the parameter values: $\alpha=0.1,~~\beta=0.2,~~\gamma=0.1,~~b=0.84$.  Second one is not a normally hyperbolic solution in phase space and from numerical investigation we observe that the set is unstable in nature (see in fig. (\ref{C2 unstable})). Therefore, it is not cosmologically interested in late time.
Cosmological evolutions of effective equation of state parameter ($\omega_{eff}$), and deceleration parameter $q$ are displayed in fig. (\ref{Hubble_evolution}) for the model for $\alpha=0.01,~~\beta=0.6,~~\gamma=0.1$ and $b=0.84$ exhibiting the phase transition from deceleration to acceleration and it will evolve in accelerating phase in future.\\

Next, from the autonomous system  (\ref{autonomous1}), we extracted seven critical points, among which all are hyperbolic in nature and the linear stability theory is sufficient to prove their stability in phase space. The cosmological features of the critical points are presented below by characterising different phases of the evolution: \\

1.{\it Radiation dominated solutions}\\

The point $P_1$ corresponds to a complete radiation dominated solution ($\Omega_{r}=1$) representing the decelerated ($q=1$) universe evolving in radiation era ($\omega_{eff}=\frac{1}{3}$). The restriction $\beta=\gamma$ will force the point $P_1$ to mimic the early evolution of the universe (because all the eigenvalues have positive real parts). Figs. \ref{Event_3D} and (\ref{Event_3D1}) show that the point $P_1$ is radiation dominated early decelerating phase. Note that the radiation dominated solution is also realized by the critical point $P_3$ when $b\longrightarrow 0$. For this case, the point $P_3$ is almost similar to that of point $P_1$. Therefore, early radiation dominated decelerating universe is obtained from our model. \\

2.{\it Matter dominated solutions}\\

Completely DM dominated ($\Omega_{m}=1$) solution represented by point $P_4$ corresponds to decelerated ($\omega_{eff}>-\frac{1}{3}$) saddle like intermediate phase of the universe when $\gamma>\beta-\frac{1}{3}$ while the point represents dust dominated decelerated phase ($\omega_{eff}=0$, $q=\frac{1}{2}$) for $\beta=\gamma$. In this case, the evolution signifies as intermediate phase of the universe (because the point is saddle in phase space). This phenomenon is shown in Figs. \ref{Event_3D} and (\ref{Event_3D1}). A similar result can be found in the point $P_5$. However, the point behaves as accelerated universe ($q=-1$) in late-time only for $\beta<\gamma+1$ (because all the eigenvalues have negative real parts). Therefore, the point is unphysical. It is worth mentioning that for restriction $\beta=\gamma$  and for limiting values of parameter $b$, the scaling solution $P_6$ will become DM dominated. In particular,  $b\longrightarrow 1$ shows that the point is DM dominated ($\Omega_{m}\approx \frac{3}{4}$, $\Omega_{d}\approx \frac{1}{4}$) decelerated evolution representing the intermediate phase of the universe (because the point is saddle like solution in phase space). On the other hand, for $b\longrightarrow 0$, the point will become DM dominated ($\Omega_{m}\approx 1,~ \Omega_{d}\approx 0$) solution evolving in dust era ($\omega_{eff}\approx0,~q\approx \frac{1}{2}$) corresponding to a decelerated transient phase of the universe (because the point is saddle in phase space). The fig.(\ref{Event_3D1}) for $\alpha=0.1,~~\beta=0.1,~~\gamma=0.1,~~b=0.93$ exhibits that the point $P_6$ is DM dominated decelerated universe signifying the transient phase. Finally, when $b\longrightarrow 0$ the DM dominated solutions will have the same nature of $P_5$.\\

3.{\it HDE dominated solutions}\\

A complete HDE dominated solution is represented by the point $P_2$. Depending upon model parameters ($b$, $\beta$ and $\gamma$), it exhibits cosmological viable characteristics. It is discussed in details in section \ref{event phase space analysis}. Late-time accelerated expansion is observed only in phantom era. However, for a particular choice of parameter: $\beta=\gamma$, and for $b=1$, the point will exhibit as late time accelerated de Sitter like solution ($\Omega_{d}=1,~\Omega_{m}=0,~\Omega_{r}=0,~\omega_{eff}=-1,~q=-1$)).  
Figs. \ref{Event_3D}, (\ref{Event_3D1}) and (\ref{P6_3D}) exhibit that the point $P_2$ is late-time accelerated attractor in HDE dominated era.

Another point $P_7$ for $\beta=\gamma$ and for $b\longrightarrow 1$ will become HDE dominated solution and it will have one dimensional non-empty stable sub manifold corresponding to negative eigenvalue ($\lambda_1=-4,~\lambda_2 \approx 0, \lambda_3\approx 0$). Then the point can be a de Sitter solution in phase space which is accelerating $\Omega_{d}\approx1,~ \Omega_{m}\approx 0, \omega_{eff} =-1,~q=-1$ where the HDE behaves as cosmological constant $\omega_{d}\approx -1$.
In fig. \ref{Event_3D} the point $P_7$ is HDE dominated accelerated solution. However, the point is not late-time solution here, rather, it is saddle. So, it is not interested physically. Similar result is shown in the fig. (\ref{P6_3D}). \\

4.{\it Scaling solutions}\\

The DM $\sim$ DE scaling solution represented by the point $P_6$ which describes the late-time evolution of the universe with constant ratio of energy densities of both DE and DM. They scale similarly in their evolution, so, the solution can solve the coincidence problem. The solution also supports the present observational data. In fig. \ref{P6_3D}, it is shown that the scaling solution $P_6$ corresponds to late-time accelerated universe attracted in phantom era. Also, the result is well agreement with the observational data:  $\Omega_{m}\approx 0.30,~\Omega_{d}\approx 0.70,~\Omega_{r}=0,~\omega_{eff}\approx-1.019,~q\approx-1.029$. Here, the HDE behaves as phantom fluid $\omega_{d}\approx-1.45$. \\

We can summarize the above results as: the model of HDE employing the Hubble horizon as IR cutoff cannot give the matter dominated intermediate phase of the universe. Therefore, a complete cosmic scenario cannot be obtained from the model when creation rate is taken as $\Gamma=3\alpha H_0+3\beta H$.  

On the other hand, the model with the future event horizon as IR cutoff and with same creation rate can provide cosmologically viable solutions from dynamical systems perspectives. We obtained a sequence of critical points showing a complete evolution of the universe. It starts from the point $P_1$ representing the early radiation dominated decelerated phase to the point $P_6$ describing intermediate decelerated matter (DM) dominated phase and finally, it enters to the point $P_2$ showing the HDE dominated late-time accelerated evolution. The figures \ref{Event_3D} and (\ref{Event_3D1}) show these scenarios clearly.
In fig. \ref{Event_3D} for parameters ($\alpha,~\beta,~\gamma,~b$)=($0.1,~0.2,~0.1,~0.93$), it is shown that $P_1$ with co-ordinate ($0,~0,~0$) is source with $\Omega_{r}=1,~\omega_{eff}=\frac{1}{3},~q=1$ showing that the point is radiation dominated decelerated solution whereas point  $P_6$ with coordinate ($0.895,~0.105,~0$) is a saddle with $\Omega_{m}\approx 0.9,~\Omega_{d}\approx 0.1,~\Omega_{r}=0,~\omega_{eff}\approx -0.1,~q\approx 0.3$ and the point $P_2$ with co-ordinate ($0,~1,~0$) with the cosmological parameters: $\Omega_{m}=0,~\Omega_{d}=1,~\Omega_{r}=0,~\omega_{eff}\approx -1.05,~q\approx -1.08$. Note that in this figure, $P_4$ also can mimic as matter (DM) dominated decelerated evolution. For this case, $P_4$($1,~0,~0$): $\Omega_{m}=1,~\Omega_{d}=0,~\Omega_{r}=0,~\omega_{eff}=-0.1,~q=0.35$. Therefore, the evolution will follow either $P_1 \longrightarrow P_6 \longrightarrow P_2$ or, $P_1 \longrightarrow P_4 \longrightarrow P_2$. A similar scenario is exhibited in the fig. (\ref{P6_3D}) for the case $\beta=\gamma$. The evolution of the cosmological parameters in fig. \ref{Event_evolution} for $\alpha=0.1,~~\beta=0.2,~~\gamma=0.1,~~b=0.93$ shows that the universe started evolving from radiation dominated era in past and passing through a long time in matter dominated era and it is at present supporting the observational data and finally it is attracted in phantom era. 
Therefore, from the phase space analysis of the model, one can conclude that several cosmological scenarios can be achieved, such as the radiation dominated early epoch, matter dominated intermediate phase and the late-time DE dominated epoch which provide viability of the model. This type of solution can also be found in \cite{Chatzarakis:2020} where phase space analysis was performed in frame of modified gravity theory. The work has also shown several cosmological scenarios like de Sitter expansion phase, radiation dominated phase and matter dominated epoch etc. The most important result like our model is existence of a heteroclinic orbit which drives the dynamical system to the stable critical point.
\\

It is worthy to note that the system (\ref{autonomous}) has only one stable critical point $C_1$ which implies its global stability as its domain of attraction is the entire space in $R^3$. On the other hand, the system (\ref{autonomous1}) admits the critical point $P_2$ as globally asymptotically stable attractor (see fig. (\ref{global stable})). The values of free parameters are: $b=0.33$, $\alpha=0.1$, $\beta=0.01$, $\gamma=1.37$ to plot the figure. For this parameters choices the fig. (\ref{P6_3D}) shows that the point $P_2$ and $P_6$ are locally stable whereas our study of global dynamics reveals that the point $P_2$ is stable globally. That means our models shows the ultimate fate of the universe is evolving in HDE dominated accelerated phase represented by $P_2$, though the scaling attractor $P_6$ is achieved for same parameter values.

\section{Concluding remarks}\label{conclusions}
In the background dynamics of FLRW universe, we investigated an interacting holographic dark energy (HDE) model by employing Hubble horizon and future event horizon as infrared (IR) cutoffs in context of adiabatic particle creation mechanism with creation rates $\Gamma=3\beta H$ (in model 1) and $\Gamma=3\alpha H_0 +3\beta H$ (in model 2). It was assumed that the universe is filled with radiation, dark energy and dark matter. Here, the universe is assumed to be an open thermodynamical model and is filled with radiation, dark matter and dark energy where the HDE is considered to be a candidate for DE and the pressureless dust as DM. Also, an adiabatic particle production is occurred irreversibly such that the negative creation pressure will have a significant contribution in the energy-momentum tensor in field equations. The created particles are assumed to be dark matter particles and which interact with HDE. A phenomenological choice of interaction $Q=3\gamma H \rho_{m}$ is undertaken for allowing the exchange of energy densities between the dark sectors (HDE and DM). In the model 1, we studied interacting HDE with Hubble horizon as IR cutoff in context of creation pressure $\Gamma=3\beta H$. In this case,  an analytic expression of deceleration parameter $q$, effective equation of state $\omega_{eff}$ and the Hubble parameter $H$ in terms of redshift ($z$) are found. This study shows that a phase transition from deceleration to acceleration is possible when interaction is allowed between dark sectors. For non-interacting case, there is no acceleration. From the expression of the squared speed of sound it is concluded that the model is not stable classically (see fig.(\ref{sound_evolution})).\\

Next, we have studied the interacting HDE with the same interaction term in model 2. Here, we considered the particle creation rate as $\Gamma=3\alpha H_0 +3\beta H$ and it is studied by adopting the Hubble horizon and the future event horizon as infrared cutoffs in two separate cases. In both the cases (in model 2), the governing equations are complicated and non-linear in form. Therefore, we performed dynamical systems analysis to obtain an overall qualitative information of the model. A proper set of dimensionless variables is chosen in these cases. Three dimensional autonomous systems of ODEs are constructed and the phase space analysis is performed for the model. Local stability of the critical points are found by evaluated eigenvalues of the linearized Jacobian matrix and stability of the model is found by finding the adiabatic sound speed.\\

In the first case (employing Hubble horizon as IR cutoff), we obtained two non-isolated set of critical points from the autonomous system (\ref{autonomous}). Set $C_1$ is normally hyperbolic set representing the dynamics at late-time cosmology which corresponds to accelerated universe evolving in quintessence era, or cosmological constant era, or in phantom regime which is shown in fig. (\ref{Point C1 C2}). Numerical investigations are carried out to determine the local stability of the model. From this, the set $C_2$ is unstable solution in phase space (see in fig. (\ref{C2 unstable})). So, the set $C_1$ is the only global attractor in the phase space. The set $C_2$ is not of much interested in late-times. Note that this model cannot give the matter dominated era. A phase transition from the deceleration phase to acceleration phase is obtained from this model (see in fig.(\ref{Hubble_evolution})). \\

Next, in the second case (taking future event horizon as infrared cutoff), and obtained a three dimensional autonomous system in Eqn (\ref{autonomous1}). We found seven critical points and all are hyperbolic in nature. We obtained critical points representing late-time accelerated universe dominated by the HDE. Some points exhibit the matter dominated decelerated phase having transient nature of evolution and some other points are radiation dominated solutions representing the decelerated evolution of early universe. Depending upon some parameters restrictions, the model can provide a complete cosmic scenario by some sequences of critical points.  
In fig. \ref{Event_3D} it is shown that a radiation dominated solution described by the point $P_1$ which corresponds to an early evolution of the universe. After that the universe is evolving towards a matter dominated decelerated phase, this phenomenon is observed by the points $P_6$ as well as $P_4$ in phase space. Finally, the universe enters into the late-time accelerated expansion phase and this acceleration is driven by HDE and the universe is dominated by HDE. This scenario is exhibited by the point $P_2$. Therefore, one may conclude that the sequences of points:
$P_1 \longrightarrow P_6 \longrightarrow P_2$ and 
$P_1 \longrightarrow P_4 \longrightarrow P_2$
can describe successfully the complete cosmic scenario from early radiation dominated phase to late-time HDE dominated phase connecting through a DM dominated intermediate phase. Note that the fig.(\ref{P6_3D}) also shows the same scenarios for restriction $\beta=\gamma$. Here, the point will describe the dust dominated decelerated era. This is discussed in detail in the previous section. Despite these, a DM $\sim$ HDE scaling solution represented by the point $P_6$ can give cosmological interesting characteristics at late times. The constant ratio in energy densities of DM and HDE leading to provide a possible mechanism to solve the coincidence problem. In fig. (\ref{P6_3D})  for $\alpha=0.1,~~\beta=0.01,~~\gamma=1.37,~~b=0.33$, it is shown the point with cosmological parameters:
$\Omega_{m}\approx 0.30,~\Omega_{d}\approx 0.70,~\Omega_{r}=0,~\omega_{eff}\approx-1.019,~q\approx-1.029$ will be well agreement with observational data. It is worthy to mention that the HDE behaves as phantom fluid $\omega_{d}\approx-1.45$ for this case. It is also interesting to mention that this type of solution cannot be obtained in non-interacting model. From the calculation of adiabatic sound speed for this model it is concluded that the model is stable throughout the evolution of the universe (see fig.(\ref{event_evolution})).

Finally, it is concluded that our study of interacting HDE model for Hubble horizon as IR cutoff in model 1 can give an accelerating universe with particle creation rate $\Gamma=3\beta H$ where as the model is unable to give accelerating phase for non-interacting case. Next, the model 2 for creation rate $\Gamma=3\alpha H_0+3\beta H$ can give the late-time acceleration when Hubble horizon is taken as IR cutoff for interacting HDE. However, scaling solutions cannot be obtained for this case. In the second case, when the future event horizon is taken as IR cutoff, the interacting HDE can be able to provide a cosmological viable solution. Moreover, the model is stable classically (since $C_s^2>0$). Therefore, the second case of the second model is more effective than that of first in the cosmological context. This is proved numerically also by plotting the time evolution of cosmological parameters: density parameter for HDE ($\Omega_{d}$), density parameter for DM ($\Omega_{m}$), density parameter for radiation ($\Omega_{r}$), effective equation of state parameter ($\omega_{eff}$), deceleration parameter ($q$). Fig. \ref{Event_evolution} confirms this for $\alpha=0.1,~~\beta=0.2,~~\gamma=0.1,~~b=0.93$. Here, the universe starts evolving from decelerated early radiation era to matter dominated decelerated intermediate era and finally enters to the HDE dominated accelerating phase. Phantom crossing behavior is achieved for this case and finally the universe evolves in phantom era in future. 
Finally, we can conclude that the model 2 is more important than that of model 1 from cosmological context. This study reveals that though the point $P_6$ shows physically interesting results at late-times evolving in accelerated era with constant ratio of DE and DM, which supports the present observational data and solving the coincidence problem. However, it cannot represent the global attractor. The study of global dynamics of the model exhibits that the point $P_2$ can be a global attractor showing that the HDE dominated late-time accelerated evolution which is the ultimate fate of the universe.

\section*{Acknowledgments}
This work is dedicated to Prof. Subenoy Chakraborty, Jadavpur University on the auspicious occasion of his 65th birthday. The authors Goutam Mandal and Sujay Kr. Biswas would like to thank Inter University Centre for Astronomy and Astrophysics (IUCAA), India where the work has been carried out in an academic visit under IUCAA's visitors' programme. The authors extend their heartfelt appreciation to the reviewers for their invaluable comments and insightful suggestions, which significantly enhanced and refined this paper. The author Goutam Mandal acknowledges UGC, Government of India for providing Senior Research Fellowship [Award Letter No. F.82-1/2018(SA-III)] for Ph.D.


\newpage

\end{document}